\documentclass[preprint]{aastex}
\usepackage{emulateapj5,psfig}
\usepackage{apjfonts}
\usepackage{natbib}

\newcommand{\lsun}{$L_{\sun}$}
\newcommand{\hubu}{km s$^{-1}$ Mpc$^{-1}$}
\newcommand{\etal}{{\rm et al.\/\ }}
\newcommand{\tm}{12$\mu$m}
\newcommand{\firr}{$f_{25}/f_{60}$}
\newcommand{\ftv}{$f_{25}$}
\newcommand{\fst}{$f_{60}$}

\newcommand{\frir}{$S_{20cm}/f_{60}$}

\newcommand{\ergs}{ergs s$^{-1}$} 
\newcommand{\oiii}{{\rm [O~III]}}
\newcommand{\oii}{{\rm [O~II]}}
\newcommand{\neiii}{{\rm [Ne~III]}}
\newcommand{\nev}{{\rm [Ne~V]}}
\newcommand{\hb}{H$\beta$}
\newcommand{\ha}{H$\alpha$}
\newcommand{\hxo}{$HX/[O~III]$}
\newcommand{\ohb}{[O~III]/H$\beta$}
\def\wave#1{$\lambda${#1}}

\slugcomment{Received 2002 August 14; accepted 2002 October 06}
\journalinfo{Accepted for publication in \it{The Astrophysical Journal}}

\shortauthors{TRAN}
\shorttitle{SEYFERT 2 GALAXIES}

\begin{document}

\title{The Unified Model \& Evolution of Active Galaxies:
Implications from a Spectropolarimetric Study}

\author{Hien D. Tran}
\affil{Department of Physics and Astronomy, Johns Hopkins University,
    Baltimore, MD 21218}
\email{tran@pha.jhu.edu}

\begin{abstract}
We extend the analysis presented in Tran (2001) of a
spectropolarimetric survey of the CfA and 12$\mu$m samples of Seyfert
2 galaxies (S2s). We confirm that S2s with polarized (hidden)
broad-line regions (HBLR) tend to have hotter circumnuclear dust
temperatures, show mid-IR spectra more characteristic of S1 galaxies,
and are intrinsically more luminous than non-HBLR S2s.  The level of
obscuration and circumnuclear star formation, however, appear to be
similar between HBLR and non-HBLR S2 galaxies, based on an examination
of various observational indicators.  HBLR S2s, on average, share many
similar large-scale, presumably isotropic, characteristics with
Seyfert 1 galaxies (S1s), as would be expected if the unified model is
correct, while non-HBLR S2s generally do not.  The active nuclear
engines of non-HBLR S2s then, appear to be truly weaker than HBLR S2s,
which in turn, are fully consistent with being S1s viewed from another
direction. There is also evidence that the fraction of detected HBLR
increases with radio power of the active galactic nucleus. Thus, not
all Seyfert 2 galaxies may be intrinsically similar in nature, and we
speculate that evolutionary processes may be at work.
\end{abstract}

\keywords{galaxies:active --- galaxies: Seyfert --- polarization}

\section{Introduction} \label{intro}

The discovery nearly two decades ago of polarized broad permitted
emission lines in NGC 1068 \citep{ma83,am85} demonstrated that some
Seyfert 2 (S2) galaxies were basically the same type of object as
Seyfert 1 (S1) galaxies but viewed from a different direction. Since then,
there have been plenty of other examples of polarized (hidden) broad-line 
regions seen in reflected light (HBLRs) in nearly all types of
active galactic nuclei (AGNs), ranging from the lowly LINERs (Barth,
Fillipenko, \& Moran 1999), to Seyferts
\citep{mg90,tmk92,t95,y96b,hlb97,mor00,t01,lum01}, to ultraluminous
infrared galaxies (ULIRGs; Hines et al. 1995, 1999; Goodrich et
al. 1996; Tran et al. 1999; Tran, Cohen, \& Villar-Martin 2000),
%\citep{hin95,goo96,t99,hin99,tcv00}, 
to powerful radio galaxies near and far (e.g., Antonucci 1984; Tran,
Cohen, \& Goodrich 1995; Cimatti \& di Serego Alighieri 1995; Young et
al. 1996a; Ogle et al. 1997; Tran et al. 1998; Cohen et al. 1999).
%\citep{a84,tcg95,cd95,y96a,o97,t98,c99}. 
This orientation-based unification model (UM; Antonucci 1993) may even 
be applicable to the luminous broad-absorption line quasars 
(BAL QSOs; Cohen et al. 1995; Goodrich \& Miller 1995; Hines \& Wills 1995).

Although the UM is widely accepted for many classes of AGN, especially
Seyfert galaxies where there are many fine examples, there is still no
consensus on its general applicability for {\it all} members of each
class. In fact, a number of studies have reported some
disturbing differences between S1s and S2s that appeared to be
inconsistent with the simple orientation-based UM. These include
suggestions that S2s tend to reside in hosts with enhanced star
forming activity \citep{mai95,gu98}, with higher frequency of
companions \citep{dr98,dh99}, or richer in dust features (Malkan,
Gorjian, \& Tam 1998, hereafter MGT98) compared to S1s.
Recently, Tran (2001, hereafter Paper I) presented the results of a
large spectropolarimetric survey of S2s from the CfA \citep{hb92} and
\tm~(Rush, Malkan, \& Spinoglio 1993) samples. The
main conclusion from this paper is that there appears to be a class of
S2 galaxies that are intrinsically weak and, as far as could be
determined, lack (or possess very weak) broad-line region (BLR) that
characterizes the genuine hidden S1 galaxies. This class of ``real''
S2s\footnote{Note that the use of ``pure'' S2s in Paper I and in Cid
Fernandes et al. (2001) refer to different types of Seyfert 2s. Cid
Fernandes et al used ``pure'' S2s to refer to those without a dominant
contribution from a starburst component, but not necessarily without
HBLR. In Paper I, we used the term ``pure'' S2s to refer to non-HBLR
S2s regardless of the starburst contribution.  To avoid confusion,
whenever possible we now use ``real'' S2s to refer to non-HBLR S2s in
this paper.} represents approximately half of the total currently
known S2 population.  In this paper, we extend the analysis of the
data from the survey of Paper I, compare them with S1s, and present
some additional evidence for the idea of two different types of S2s 
and its implications.  We assume $H_o$ = 75 \hubu, $q_o$ = 0 and
$\Lambda$ = 0 throughout this paper.

\section{Data \& Results} \label{obs}

The spectropolarimetric observations of the CfA and \tm~S2s were
briefly described in Paper I. All of the observations were made at 
at Lick and Palomar Observatories, except one (for F08572$+$3915) which
was obtained at Keck Observatory. These observations were made with the
main goal of searching for polarized broad \ha, which is the strongest
of the hydrogen Balmer lines. Accordingly, they were optimized for 
the red spectral region, around the wavelength of redshifted \ha. 
Observations at Lick were made with the 3-m Shane telescope and the
Kast double spectrograph \citep{ms93}, using a dichroic that
splits the light at 4600 \AA. A 600 grooves mm$^{-1}$ grating was
used on the red side, while a 600 grooves mm$^{-1}$ grism was used
on the blue side, providing a resolution of $\sim$ 6 \AA~for both sides.
The wavelength coverage was typically 3200--4500 \AA~in the blue, and 
4600--7400 \AA~in the red, both on 400$\times$1200 pixel CCDs.
At Palomar, spectropolarimetry was obtained with the double spectrograph
\citep{og82} on the 5-m Hale telescope. In combination with a
5500 \AA~dichroic, we used a 300 grooves mm$^{-1}$
grating on the blue, and 316 grooves mm$^{-1}$ grating on the red, giving
a typical wavelength coverage of 3600--5500 \AA~and 5500--8000 \AA, 
respectively, with 800$\times$800 pixel CCDs. The spectral resolution
was about 6 \AA~in the red and 8 \AA~in the blue. 
The single observation at Keck was made on 1994 October 29 (UT) with the 
polarimeter module installed on the Low Resolution Imaging Spectrometer 
(LRIS, Oke et al. 1995) at the 10-m Keck I telescope. A 300 grooves mm$^{-1}$
was used with a 2048$\times$2048 CCD detector to give a wavelength coverage
of 3900--8900 \AA, and a resolution of $\sim$ 10 \AA.   
For all observations, we employed a long slit with width ranging from
1\arcsec~(Keck) to 2\arcsec~(Palomar) or 2\farcs4 (Lick), centered on 
the nucleus and oriented mostly east-west, and in some cases near the 
parallactic angle. Data were reduced using standard VISTA procedures used 
in previous studies (see e.g., Tran 1995). 
  
As in Paper I, we will refer to galaxies classified as HII, LINERs, or
starburst galaxies as the HLS sample. There are 16 such sources and
they listed in Table~1. We will display these galaxies
in figures, but have excluded them from all statistical analyses in
this paper for Seyfert galaxies.  The total number of S2s in the
\tm~sample is 51, of which 43 have been observed either by this or
other studies. Excluding the intermediate Seyferts (i.e., S1.8, S1.9),
all 14 CfA galaxies classified as S2 by OM93 have been observed
spectropolarimetrically.
Most of the remaining eight un-observed S2s are unreachable by
telescopes employed in the survey.  The main disadvantage of the
\tm~sample is that its spectroscopic classification is much poorer
compared to that of the CfA sample, which has been further refined by
Osterbrock \& Martel (1993, OM93).  In fact, the classification of
\citet{rms93} was found to contain many misclassifications.  In our
study, we take advantage of the high signal-to-noise ratio (S/N)
spectra as a by-product of the spectropolarimetric observations and
re-classify these objects.

The main result of the survey, the presence or absence of HBLR
detection for our sample galaxies, is presented in Table~1, 
along with their most relevant X-ray, optical, IR,
and radio properties. References are given for the source of the data,
most of which have been collected from the literature. The optical
data, such as \oiii~flux and Balmer decrement, when not previously
available, have been measured directly from our spectroscopy. We note
that only one galaxy (F08572$+$3915), which belongs in the HLS class,
was observed at Keck.  Although the S/N is superb for this object, it
does not show HBLR. All other sources were observed at Lick or Palomar
Observatory. Thus to a good approximation, the detection of HBLR in
the sample S2s is probed to similar depth.  Results from the following
surveys have also been used: Miller \& Goodrich (1990); Tran, Miller,
\& Kay (1992); Young et al. (1996b); Heisler et al. (1997); Barth et
al. (1999); Moran et al. (2000); Lumsden et al. (2001). The literature
data for S1s for the combined CfA and \tm~sample are shown in Table~2.  
We have revised the original S1 table of \citet{rms93}
to include only S1 and S1.5 galaxies, excluding those classified as
S1.8 or S1.9.  In order to avoid biases, we have also removed several
highly radio luminous 3C galaxies from their S1 list, as noted in the
table.

\subsection{Discrepancy in HBLR S2 fraction between the CfA and \tm~Samples ?} \label{diff}

In paper I, it was noted that the detection rate of HBLRs is
significantly lower in the CfA sample (4/14 = 29\%) than the
\tm~sample (21/43 = 49\%), although the CfA detection rate is similar
to that reported by previous studies (e.g., Moran et al. 2000).
Because the \tm~sample selects objects in the IR, and therefore is well suited
for picking up galaxies that are dust obscured, it was suggested that
the $\approx$ 50\% HBLR detection rate may be more
representative than the lower 30--35\% suggested by the CfA sample and
other optically defined samples.  The optically selected CfA sample,
on the other hand, while avoiding the normal biases suffered by the
traditional UV-access searches by spectroscopicly identifying a
magnitude-limited sample of nearby galaxies, may have missed the more
dust obscured AGNs.

Note that in surveying the CfA sample, we chose to observe only those
classified as strict S2 by OM93. Seyfert 1.8s and 1.9s were
intentionally excluded. This was not done for the \tm~sample mainly
because no such detailed classification was available for it. Could
this have been the source of the discrepancy in the detection rate?
This is unlikely to be the case. Goodrich (1989) carried out a
spectropolarimetric survey of Seyfert 1.8s and 1.9s and found that
only three out of 12 such galaxies showed polarization, indicating that their
broad lines were due to scattered light.  If we assume a similar
detection rate for the S1.8s and S1.9s in the CfA
sample\footnote{Incidentally, two of the objects surveyed by Goodrich
(1989) is in the CfA sample (Mrk 744 and Mrk 471), both of which
turned out to show little or no polarization.}, this suggests that 
not including them in the survey cannot explain this discrepancy.

Recently, Moran et al. (2001) reported that two of the CfA S2s
identified as non-HBLRs in Paper I (NGC 5347 and NGC 5929)\footnote{NGC 5929
was also reported as a non-HBLR S2 by \citet{lum01}} were found
to show weak polarized broad lines in more sensitive Keck
observations.  If confirmed, it would bring the HBLR S2 fraction in
the CfA to 43\%, more consistent with that reported for the
\tm~sample. Thus, there may not be a discrepancy in the HBLR S2
fraction between the two samples, and 50\% remains a good
representative value for the fraction of the total S2 population that
contains powerful hidden S1 nuclei. However, to avoid mixing surveys
with different detection limits and to remain consistent our original
detection limit of 3--5m class telescopes, in the rest of the analysis
in this paper, we opted to keep these two objects in the non-HBLR
sample.  The difference caused by moving them to the HBLR
classification is small, and does not significantly alter the main
conclusions of this paper.

Since non-HBLR S2s are shown to be systematically weaker than their
HBLR counterparts (this work; Paper I; Lumsden \& Alexander 2001), it
is possible that with deeper observations, some of the non-HBLR S2s
reported in this paper (such as NGC 5347 and NGC 5929) may turn out to
show weak polarized broad lines. However, as we will show in the rest
of the paper, the majority of the non-HBLR S2s are probably ``real''
S2s that may not contain a genuine S1 nucleus. This is based mainly on the 
finding that their large-scale properties are systematically different from 
both HBLR S2s and normal S1s, showing that they cannot be the same type of
objects seen from another direction. This is in contrast to the
HBLR S2s, which are true S1 counterparts and whose properties match
those of S1s.

Although many of the HBLR S2s detected by optical spectropolarimetry
have also been observed to show broad permitted lines directly in
near-infrared spectroscopy, which presumably probes deeper through the
obscuring torus, the correspondance of BLR detection between the two
different methods is generally poor (e.g., Veilleux, Goodrich, \& Hill
1997; Lutz et al. 2002). The rate of BLR detection in S2s by direct
near-IR

%TABLE 1
\begin{figure*}[t]
\centerline{\psfig{file=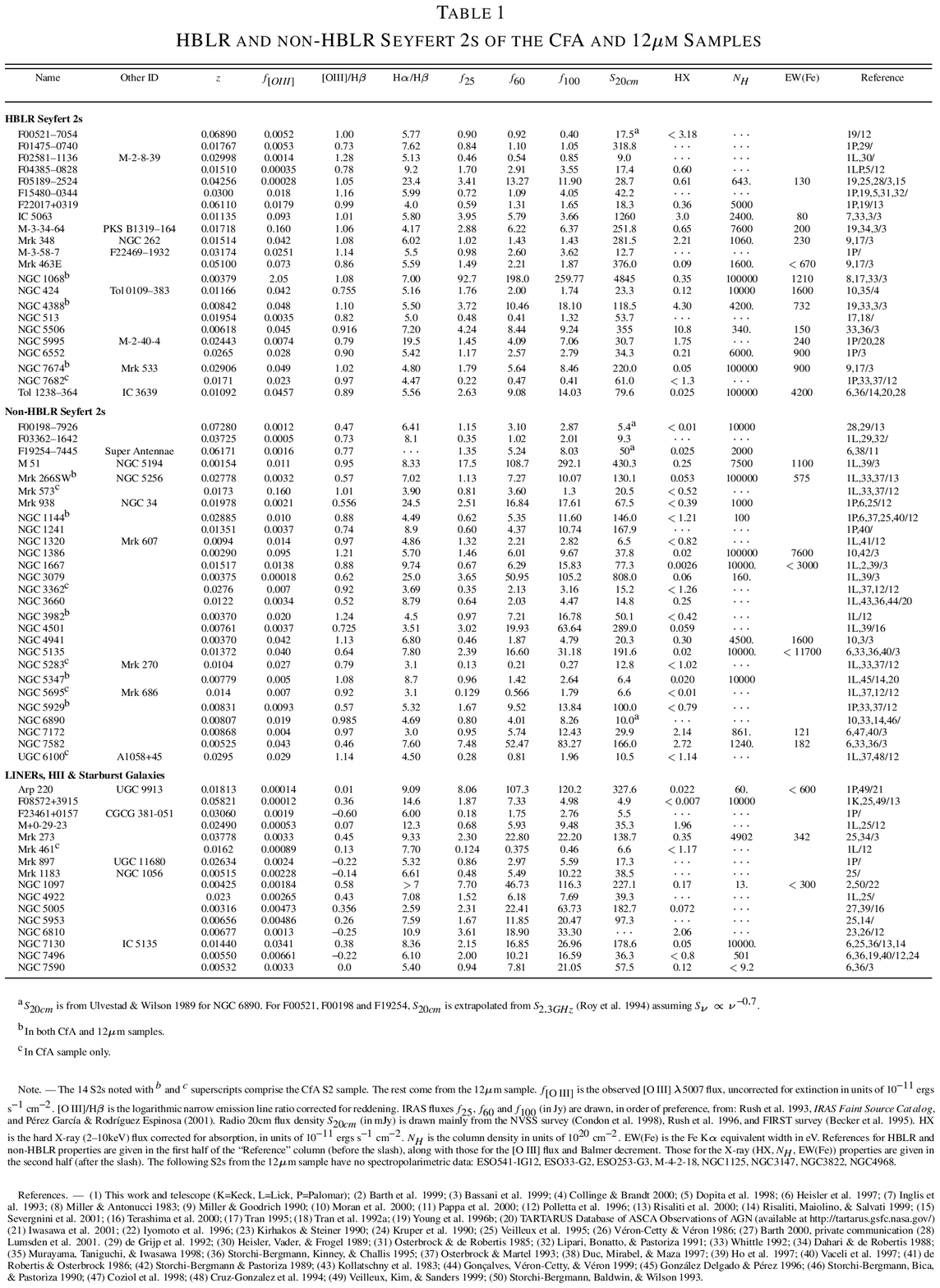,width=18.5cm,angle=0}}
\end{figure*}
\clearpage

%TABLE 2
\begin{figure*}[h]
\centerline{\psfig{file=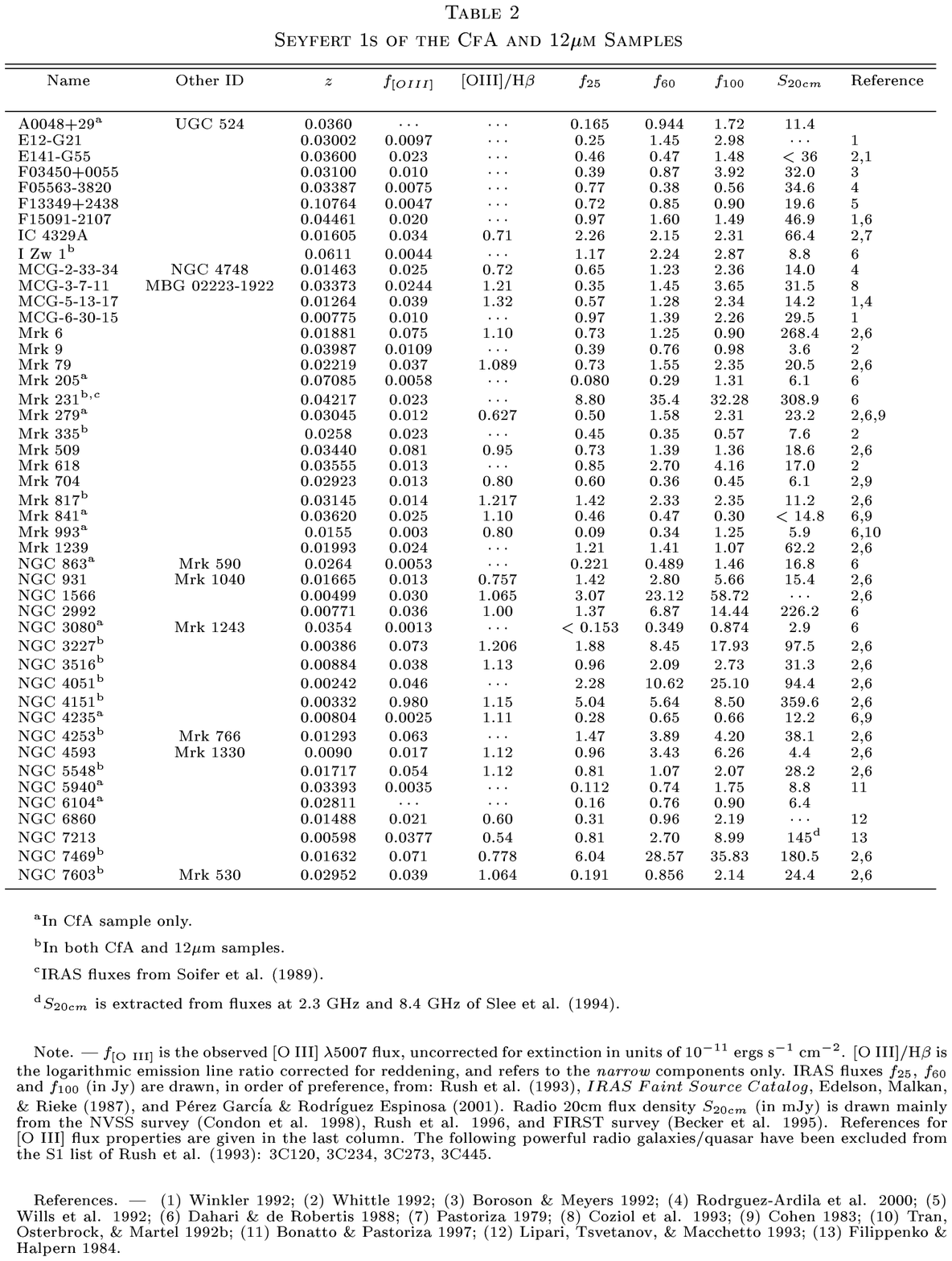,width=18.5cm,angle=0}}
\end{figure*}
\clearpage

\noindent
spectroscopy is around 25\%, somewhat lower than that by
spectropolarimetry, and many of the S2s known to have polarized
optical broad lines fail to show a corresponding near-IR broad line,
such as Br$\alpha$, in direct flux \citep{lut02}. These results imply
that in some S2s, the obscuring material is still considerably
optically thick at $\sim$ 4$\mu$m.

\subsection{Diagnostic Diagrams \& Luminosity Distributions} \label{diag}

In order to compare various properties among the HBLR, non-HBLR S2s
and S1s, we now present several diagnostic diagrams which aim to
illustrate their similarities and differences. The results of our
statistical tests are summarized in Table~3.  We will
first examine the [O III]/\hb~vs. \firr~plot shown in Figure
\ref{oiiihb}. As discussed in Paper I, these two ratios display
significant
%FIG 1
\psfig{file=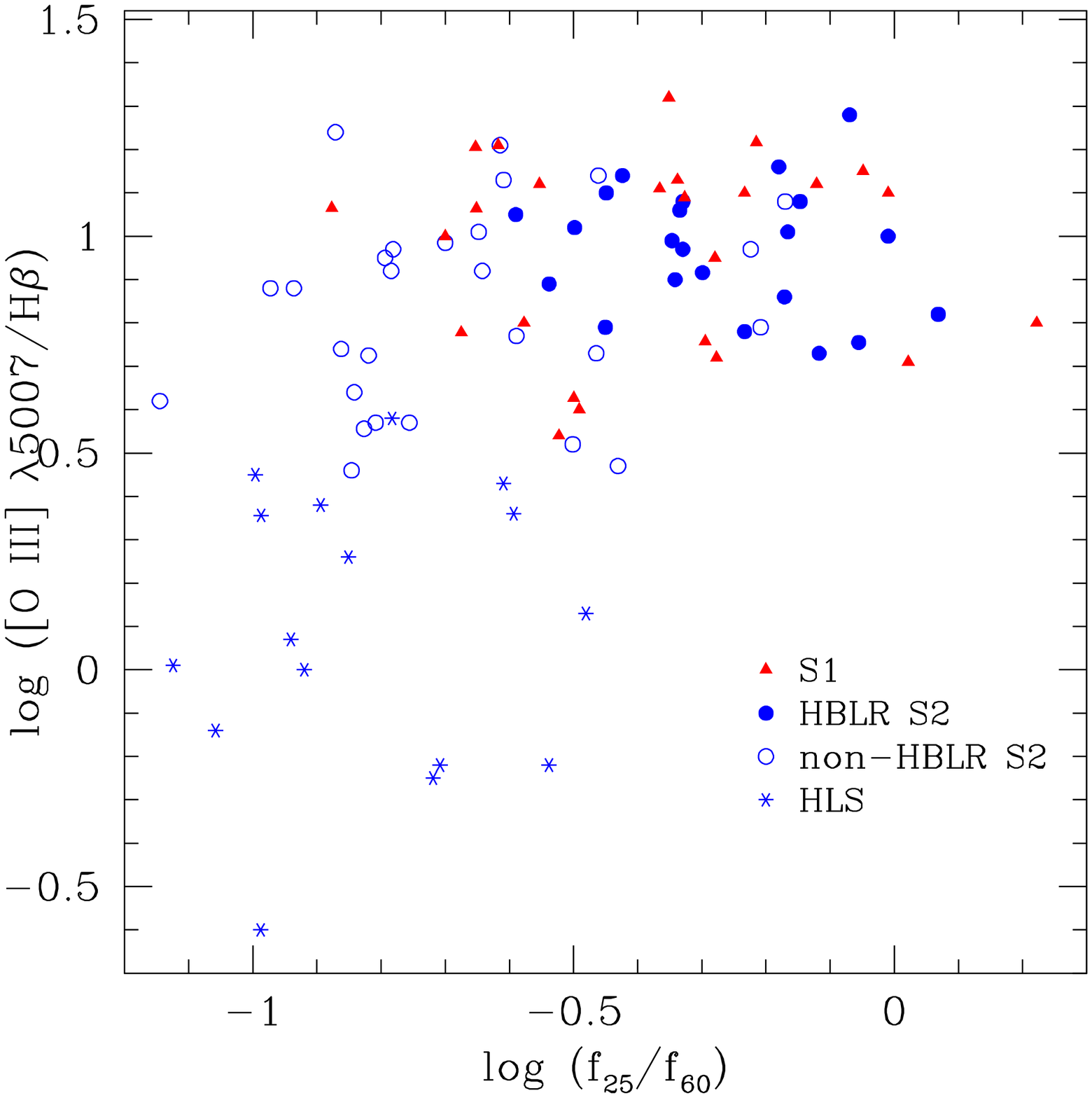,width=8.6cm}
\figcaption{Ionization measure \oiii/\hb~versus IR color \firr~for the
CfA and \tm~samples. Seyfert 1 galaxies are shown as solid triangle,
HBLR S2s as solid dots, and non-HBLR S2s as open circles.  Asterisks
denote HII/LINER/SB (HLS) galaxies, all of which have no HBLRs.
\label{oiiihb}}
\vskip 0.3cm

\noindent
differences between the two S2 classes.  What remains to
be determined is how they compare to the S1 population.  In
considering the S1s, we must keep in mind that the \ohb~ratio refers
only to the narrow-line component. Thus, strictly speaking, only type
1 Seyferts that show prominent \hb~narrow component, such as Seyfert
1.5s should be considered. Thus, we have gathered the relevant data
from the literature for S1.5s, which are listed in Table~2
and plotted in Figure \ref{oiiihb} along with those for both HBLR and
non-HBLR S2s. The distributions of [O III]/\hb~and \firr~as a function of
Seyfert types are shown in Figures \ref{lohbhist} and
\ref{irchist}.  A visual examination quickly confirms that indeed
S1.5s do tend to show similar \ohb~and \firr~ratios to HBLR S2s,
suggesting that they are intrinsically the same type of object.  A
formal K-S test shows that statistically the mean \ohb~and \firr~for
the S1.5s are not significantly different from those shown by the
HBLR S2s (Table~3)\footnote{We shall adopt the traditional
view that a test result with $p_{null} \leq 5\%$ is considered to be
significant.}.  Compared to non-HBLR S2s, however, the S1.5s display
significantly higher values of these quantities.  Thus, not only are
non-HBLR S2s different from HBLR S2s, the latter appears to be similar
to normal S1s. These results provide 

%FIG 2
\psfig{file=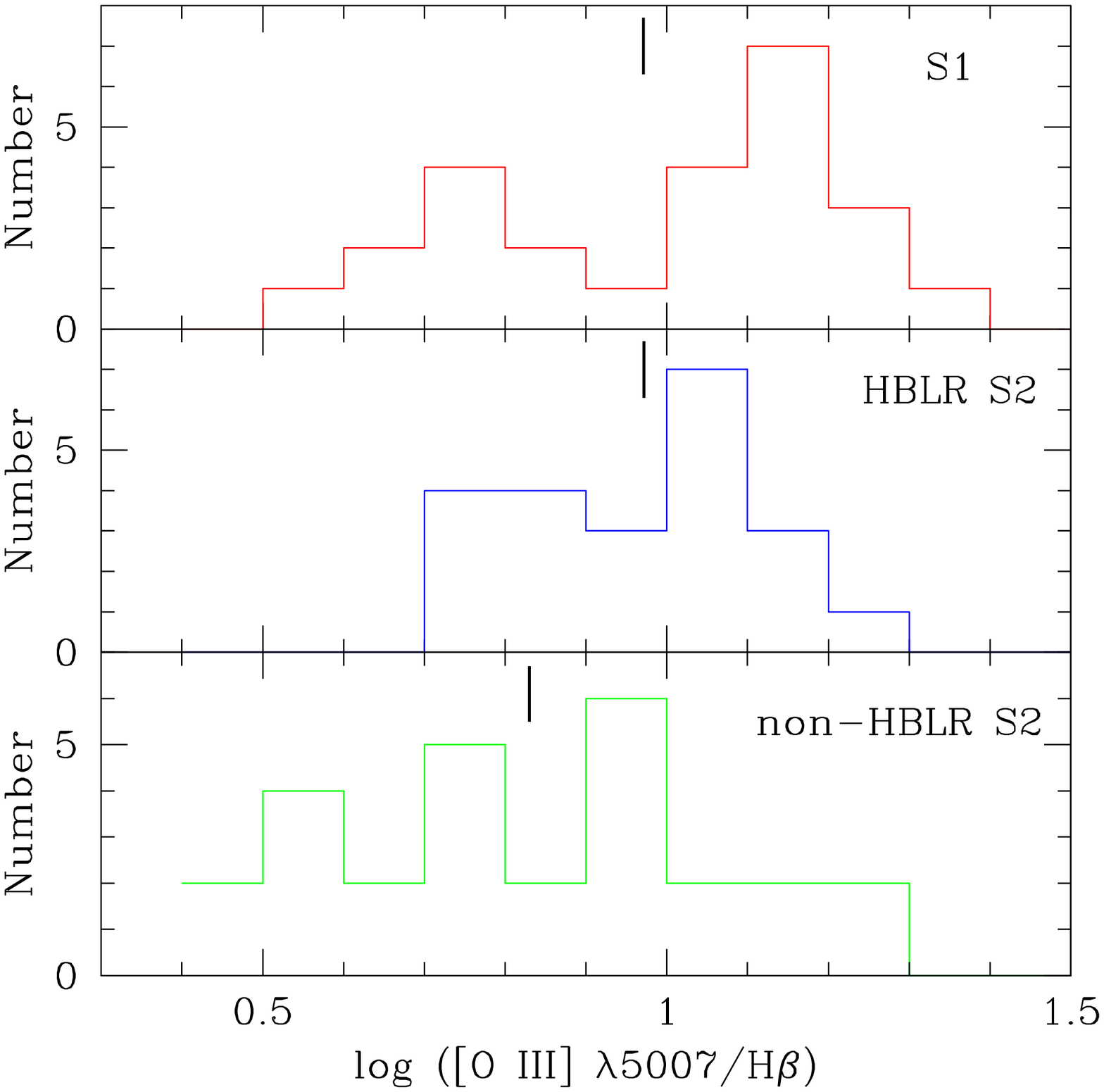,width=8.6cm}
\figcaption{The distribution of \oiii/\hb~ratio for the combined CfA and
\tm~samples of Seyfert 1 galaxies ({\it top}), HBLR S2s ({\it
middle}), and non-HBLR S2s ({\it bottom}). The vertical tickmark in each 
panel denotes the mean of each distribution. The S1s show similar
distribution to HBLR S2s, both of which are significantly different
from non-HBLR S2s.
\label{lohbhist}}
\vskip 0.3cm

\noindent
additional support for the
concept put forward in Paper I that the two types of S2s are
different, with one being truly obscured S1s, and the other having
much less powerful central AGN.

Note in Figure \ref{oiiihb} that the lower right corner can be populated by
HBLRs, indicating that not all S2s with low \ohb~ratio are necessarily
non-HBLRs. This could be explained as a result of a combination of
obscuration of the NLR and mixing of starburst
and AGN components (e.g., see Hill et al. 2001;

%FIG 3
\psfig{file=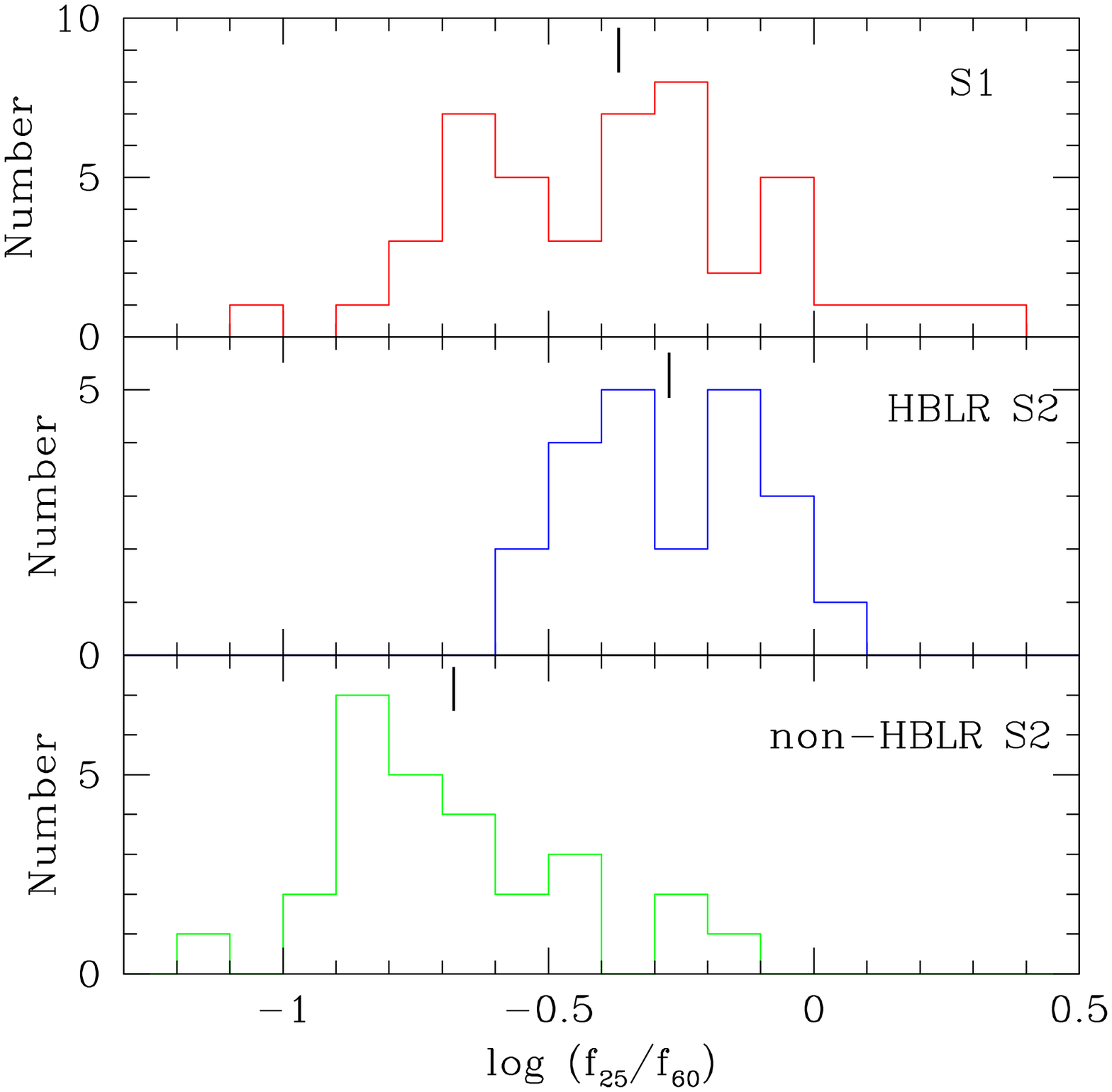,width=8.6cm}
\figcaption{The distributions of the IR color ratio \firr, arranged as in
Figure \ref{lohbhist}. The S1s show similar distribution to HBLR S2s,
both of which are significantly different from non-HBLR S2s.
\label{irchist}}
\vskip 0.2cm

%TABLE 3
\begin{figure*}[t]
\centerline{\psfig{file=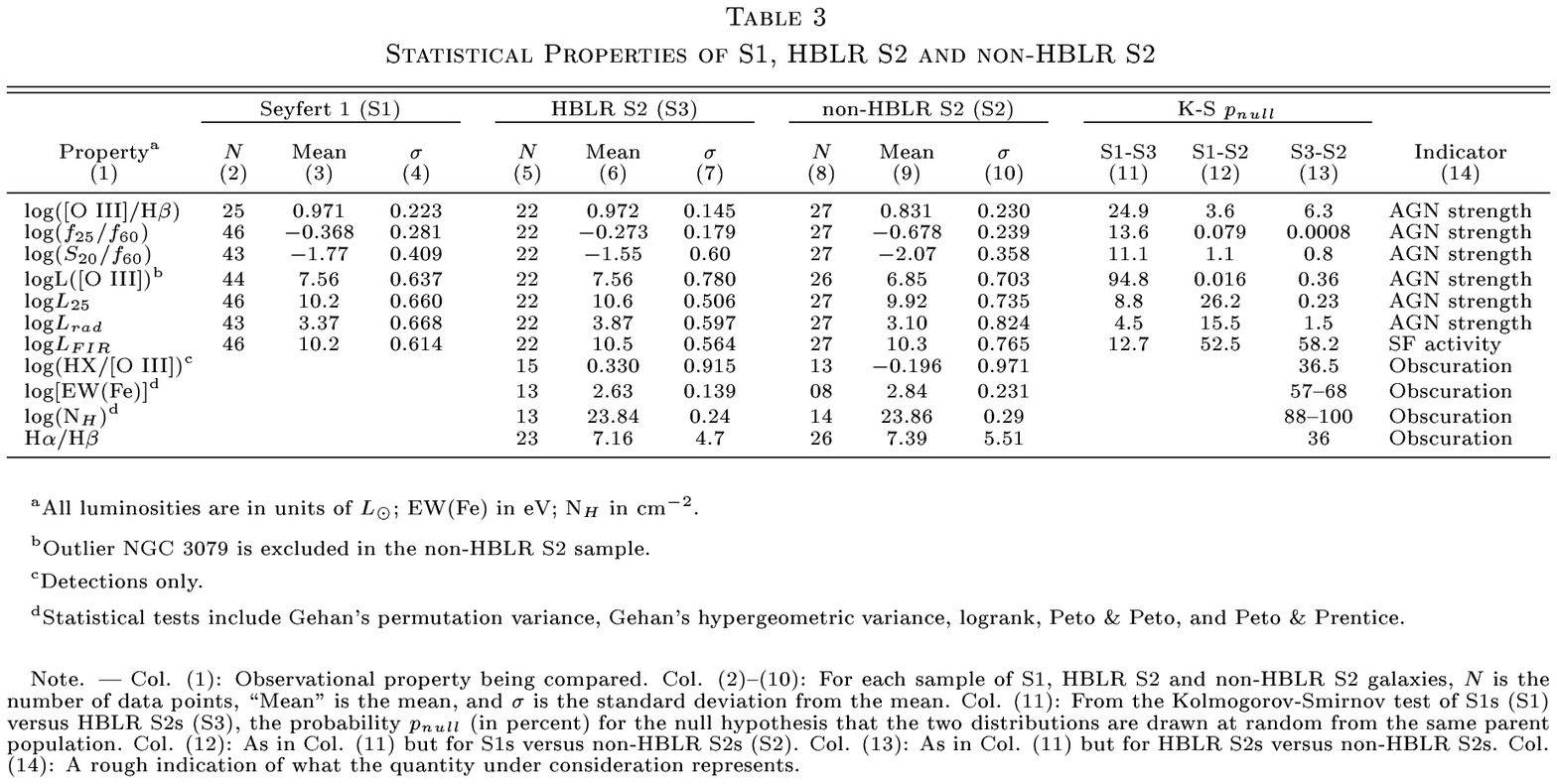,width=18.5cm,angle=0}}
\end{figure*}

\noindent
Levenson et al. 2001a). The lower line ratio could also arise in part from
partial obscuration either by the obscuring torus or dust of the
higher ionization lines close to the nucleus. Evidence for such
obscuration has come from the observation of stratification of the
polarization of narrow emission lines, in the sense that higher
ionization lines are higher polarization \citep{bfm99,tcv00}. The
\firr~ratio, however, is not significantly affected by the
obscuration, maintaining an essentially warm color. Thus, HBLR S2s
lying in this region are likely to be dusty S2 galaxies with a mixed
starburst component, having extended dust lanes that could obscure
much of the high-ionization optical emission close to the nucleus.

%FIG 4
\psfig{file=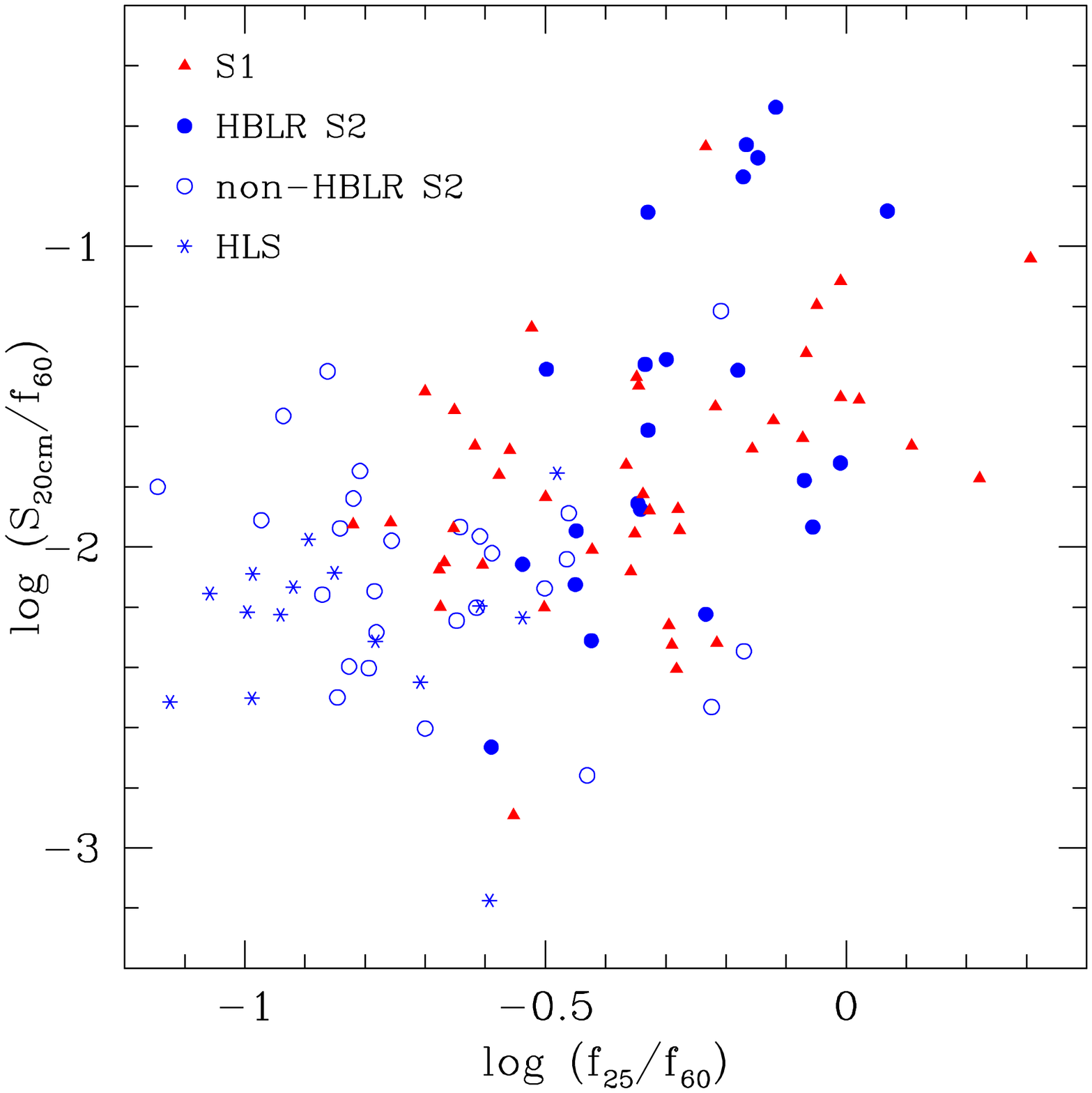,width=8.6cm}
\figcaption{20cm radio flux density $S_{20cm}$, normalized by the FIR
flux \fst, which is dominated by star formation in the host galaxy, as
a function of IR color \firr~for the CfA and \tm~samples.  Symbols are
as in Figure \ref{oiiihb}.
\label{radirc}}
\vskip 0.2cm

We turn next to the diagram of \frir~vs. \firr, which has been shown
in Paper I to display a markedly clear segregation between S2 types.
In Figure \ref{radirc} we added in the S1 data. Again, the S1s show a
strong tendency to lie among the HBLR S2s and to avoid the region
inhabited by non-HBLR S2s.  The distributions of \frir~(Figure
\ref{lrirhist}) confirm these behaviors, and K-S tests show that S1s
and HBLR S2 are statistically alike, with both being significantly
different from the non-HBLR S2s (Table~3). Note that most
of the radio data used come from the 1.5GHz NVSS survey. Higher
resolution data for both samples are available at 8.3GHz (Thean et
al. 2000, 2001; Kukula et al. 1995), but the shorter wavelength radio 
emission may have a

%FIG 5
\psfig{file=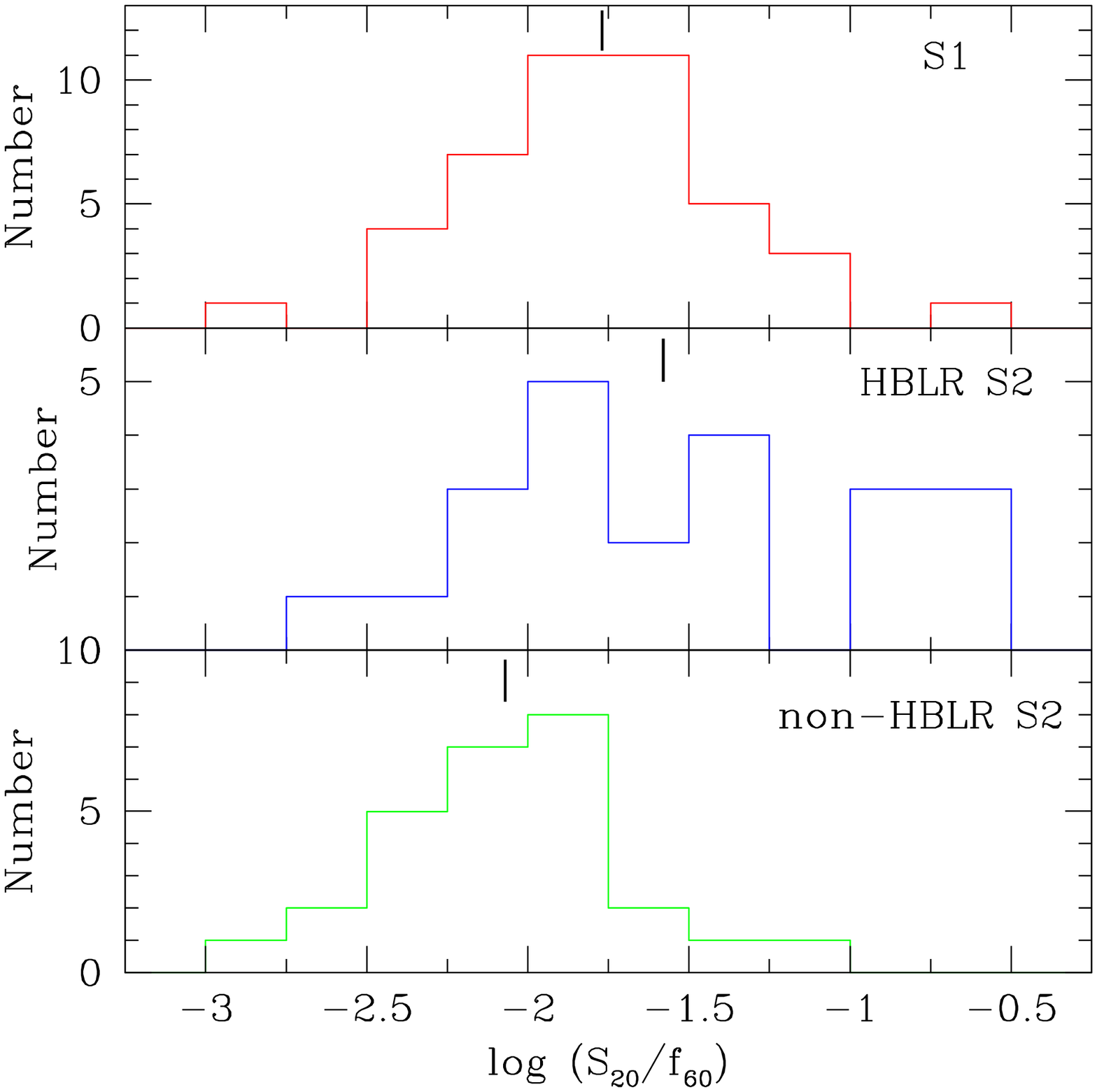,width=8.6cm}
\figcaption{The distributions of the \frir~ratio, arranged as in Figure
\ref{lohbhist}.The S1s show a similar distribution to HBLR S2s, both
of which are significantly stronger than non-HBLR S2s.
\label{lrirhist}}
\vskip 0.2cm

\noindent
higher contribution from star formation in the host. 
In addition, since we are considering the ratio of radio flux to
$IRAS$ far-IR flux, which is more comparable to NVSS resolution,
the NVSS data would be more appropriate for this purpose. By
normalizing to the far-IR, which is dominated by extended star formation, 
much of the non-nuclear radio emission has been accounted for, and the 
nuclear contribution has, in effect, been isolated. Using their 
high-resolution 8.3GHz data, \citet{the01} have also done a comparison 
of radio power between S1s and the two S2 subtypes. Their results
are basically consistent with ours in finding that the HBLRs are more 
powerful in the radio than non-HBLRs (see below), suggesting that 
the difference in resolution in the radio data does not have any significant
impact on the results.

Paper I has shown that the mean hard X-ray column density as well as
the Balmer decrement between non-HBLR and HBLR S2s are not
significantly different, indicating their similarity in nuclear
obscuration.  To further examine if obscuration plays a role in the
detection/visibility of HBLR in S2s, we wish to explore other
potential measures of obscuration.  HX luminosity is reflective of the
strength of the AGN, but it is also sensitive to obscuration.  In
order to isolate the effect of obscuration alone, we consider the
ratio \hxo.  Since the \oiii~strength is largely a measure of the
strength of the AGN, by taking the ratio with \oiii~we have
effectively ``divided out'' the AGN component, leaving essentially a
measure of obscuration. This ratio, called ``T'' in \citet{b99}, has
been shown to be a good indicator of obscuration \citep{b99,pgs01}.
In particular, it is anticorrelated with both the column density $N_H$
and the K$\alpha$ iron line equivalent width EW(Fe). In Figure
\ref{12mewt}, we plot $N_H$ and EW(Fe) against \hxo~for our sample of
\tm~S2 galaxies.  Both the $HX$ and \oiii~fluxes have been corrected
for obscuration, or extinction. We confirm that there appears to be a good

%FIG 6
\psfig{file=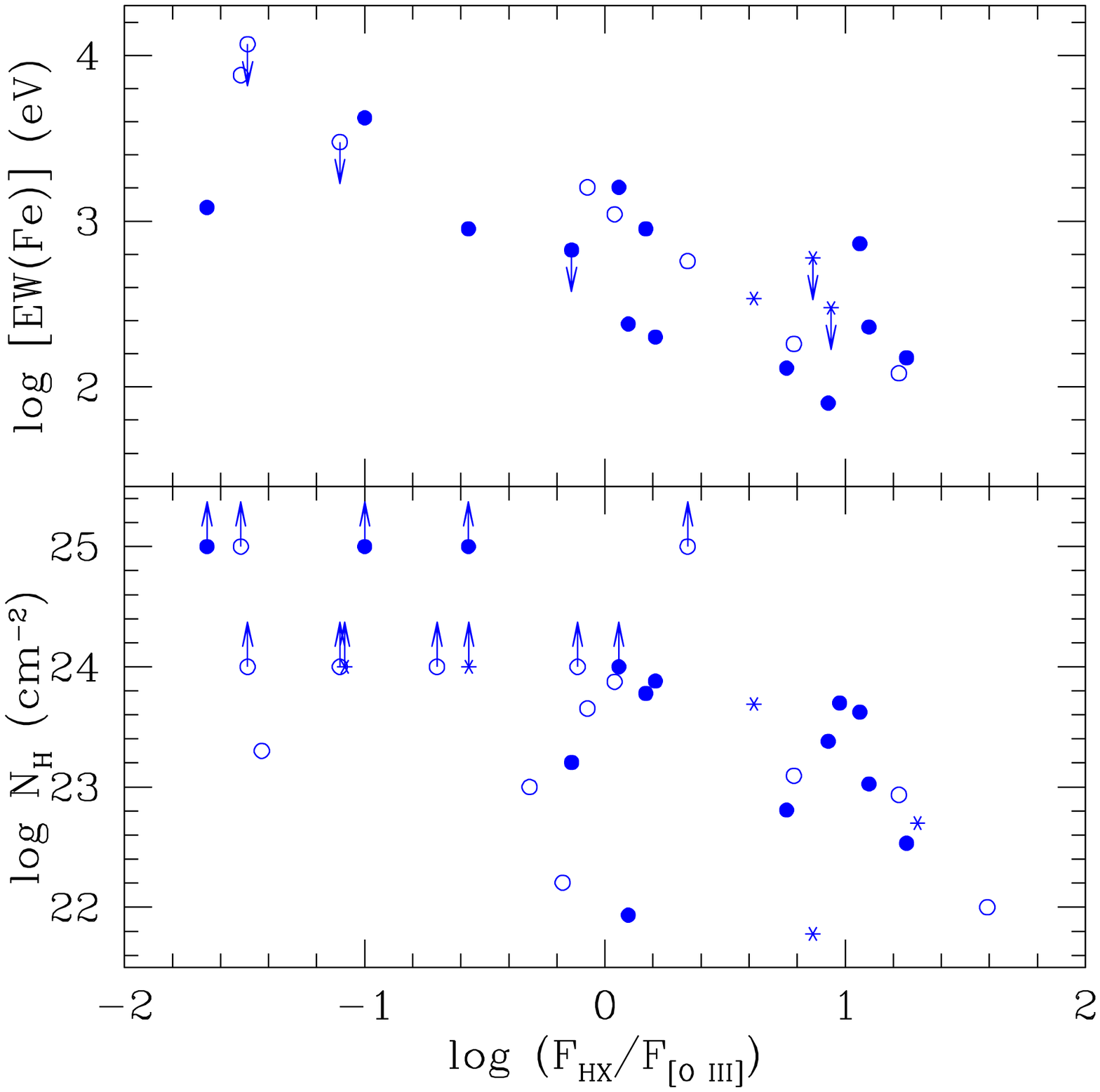,width=8.6cm}
\figcaption{The \hxo~ratio versus absorbing column density $N_H$ and
equivalent width of the K$\alpha$ Fe emission line for S2s and HLS
galaxies in the \tm~sample.  Symbols are as in Figure \ref{oiiihb};
arrows denote upper or lower limits. There is a good anticorrelation between
\hxo~and both $N_H$ and EW(Fe), indicating that these parameters can
be used as measures of the nuclear obscuration.
\label{12mewt}}
\vskip 0.2cm

\noindent
anticorrelation between the \hxo~ratio and both $N_H$ and EW(Fe),
and that these quantities can be used as probes of the obscuration to
the center of the active nucleus.

As shown in Figure \ref{hxohist}, the distribution of \hxo~appears to
be very similar between the two classes of HBLR and non-HBLR S2s.  A
K-S test of only the detected sources (no detection limits) shows that
this ratio is virtually identical between HBLR and non-HBLR S2s
($p_{null} = 36.5\%$).  Taking into account censored (i.e., upper HX
limits) data\footnote{Using the ASURV package in IRAF.}, however, it
appears to show that non-HBLR S2s may have a significantly
($p_{null}=3\%-9\%$) higher obscuration than HBLR S2s. Better X-ray
detections, perhaps with $Chandra$ or $XMM$, may be able to confirm
this difference. Turning to the EW(Fe), statistical tests also confirm
that there is virtually no difference between the samples of 13 HBLR
S2s and eight non-HBLR S2s ($p_{null}=57\%-68\%$) with available data,
which are plotted in Figure \ref{12mewt}. Therefore, after examining
various possible observational indicators for it, we conclude that the
level of obscuration is largely indistinguishable between the two
types of S2s, confirming the suggestion of Paper I that it does not
play a great role in the detectability/visibility of HBLR.

%FIG 7
\psfig{file=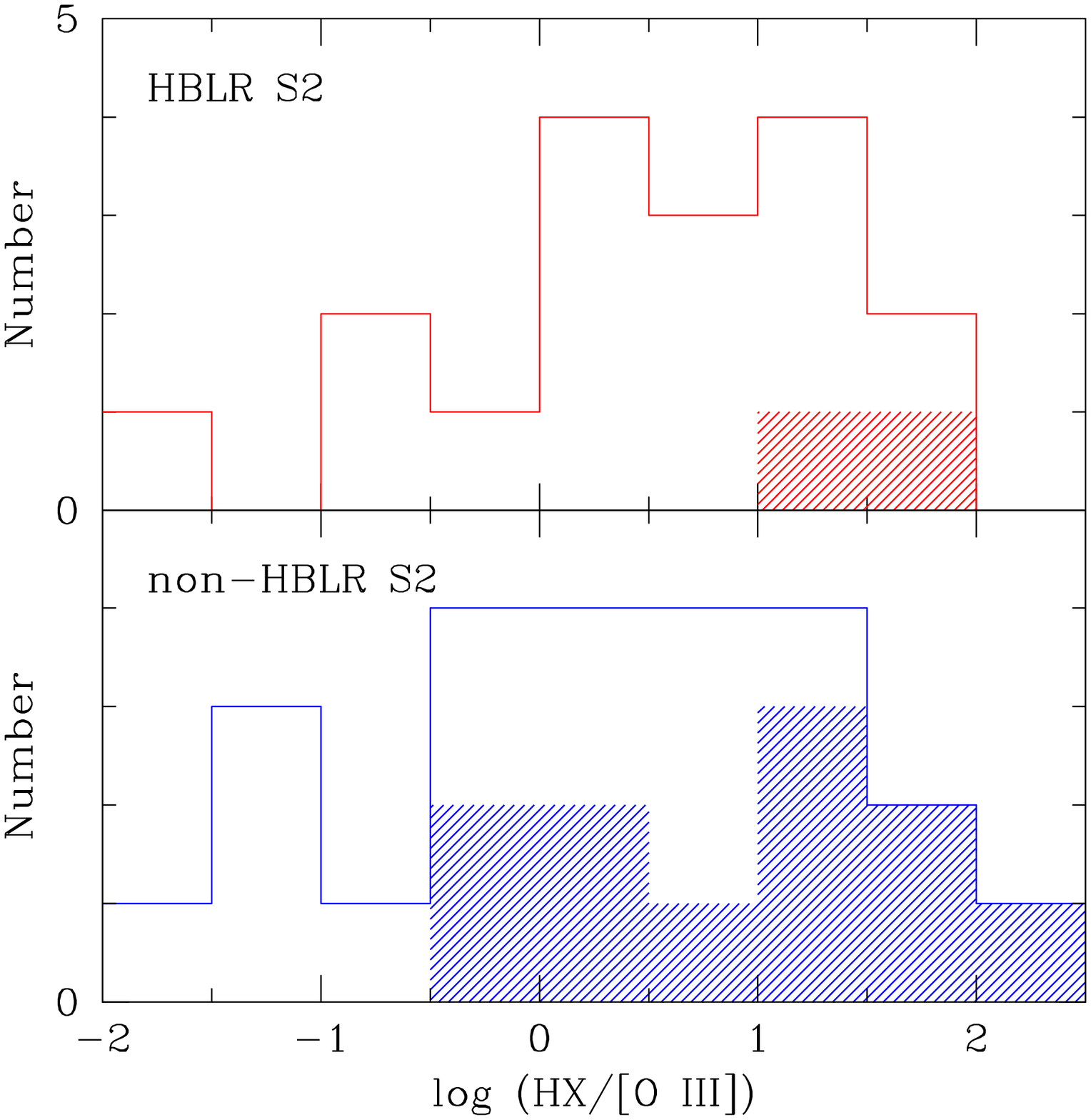,width=8.9cm}
\figcaption{Distribution of the \hxo~ratio for HBLR S2s ({\it top}),
and non-HBLR S2s ({\it bottom}) in the CfA and \tm~samples. Shaded
areas denote upper limits.  Excluding detection limits, there is no
significant difference in the mean \hxo~between HBLR and non-HBLR S2s,
indicating that non-HBLR S2s are not any more obscured than HBLR
S2s. Including the limits results in a modest significance in the
difference between the two distributions.
\label{hxohist}}
\vskip 0.3cm

We next consider HX vs. \oiii~luminosities, shown in Figure \ref{hxoiii}.  
The diagram can be divided into four quadrants with the dividing lines
roughly at L(HX) = $10^{42.4}$ \ergs~and L(\oiii) = $10^{41.5}$ \ergs.
There is a good positive correlation between these two quantities, as
would be expected, but there is also considerable scatter, which could
arise from two sources: variability in the intrinsic X-ray flux or in
absorbing column density (e.g., Risaliti, Elvis \& Nicastro 2002;
Smith, Georgantopoulos, \& Warwick 2001).
In the upper right quadrant lie mainly the HBLRs; these
are the strong AGNs with genuine hidden S1 

%FIG 8
\psfig{file=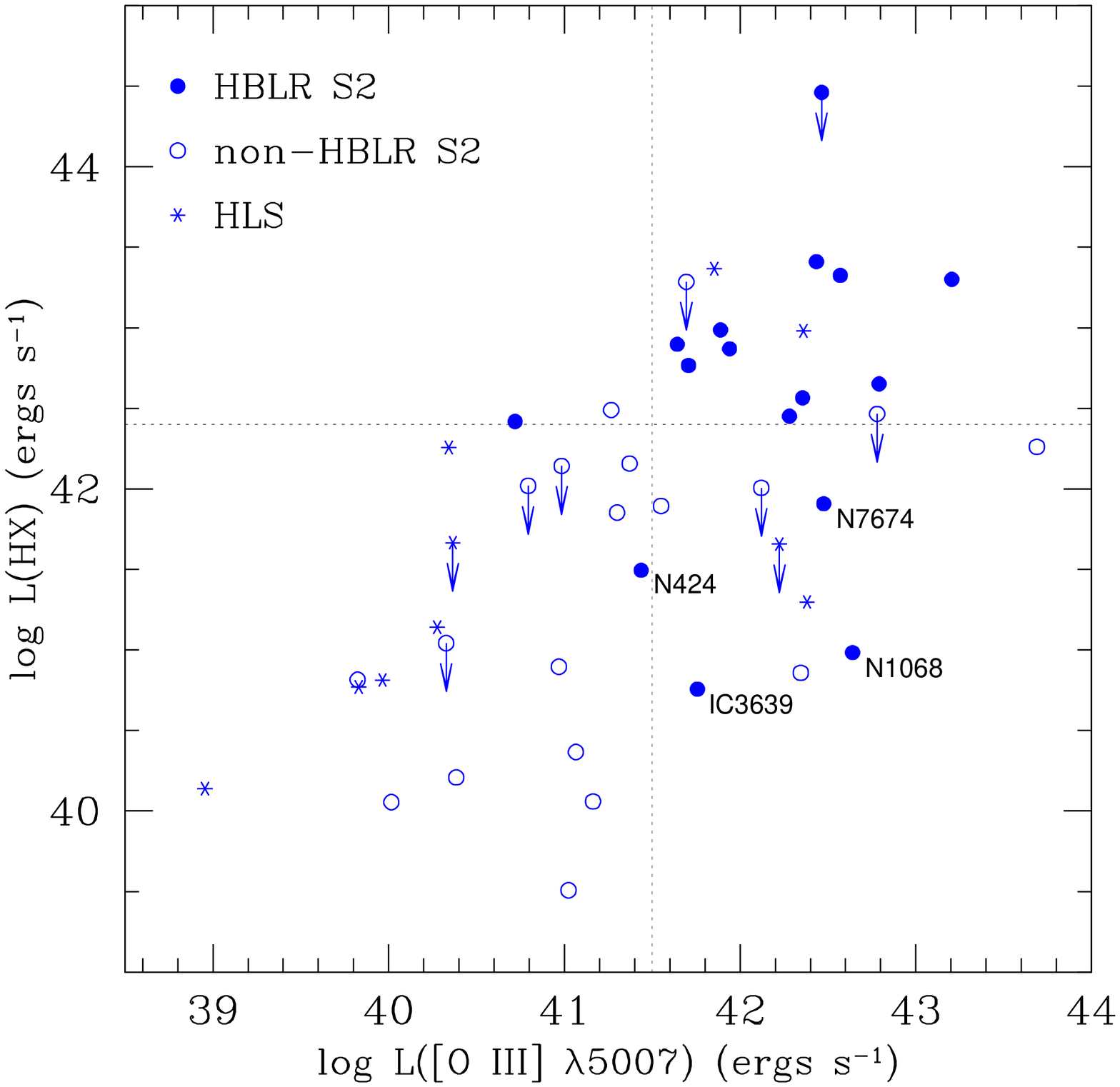,width=8.6cm}
\figcaption{Hard X-ray luminosity versus optical \oiii~\wave 5007
luminosity for CfA and \tm~Seyfert 2 galaxies.  Symbols are as in
Figure \ref{oiiihb}. The dotted lines show the rough division between
HBLR and non-HBLR S2s.  Aside from the effect of absorption on the
X-ray strength, there is a good correlation between L(HX) and
L(\oiii), with the HBLR S2s being stronger.
\label{hxoiii}}
\vskip 0.2cm

\noindent
nuclei. The lower right quadrant is occupied by similarly
powerful AGN with HBLRs, but these
suffer from high obscuration; they are the so-called Compton-thick
AGNs. All of the four labeled HBLR occupants in this quadrant (IC3639, 
N424, N1068, N7674) have $N_H > 10^{24}$ cm$^{-2}$. 
The vast majority of the non-HBLRs lie in the lower left
quadrant; these are the intrinsically weak AGNs. The lack of objects 
in the upper left quadrant is real: hard X-ray luminous AGNs 
are not expected to show 

%FIG 9
\psfig{file=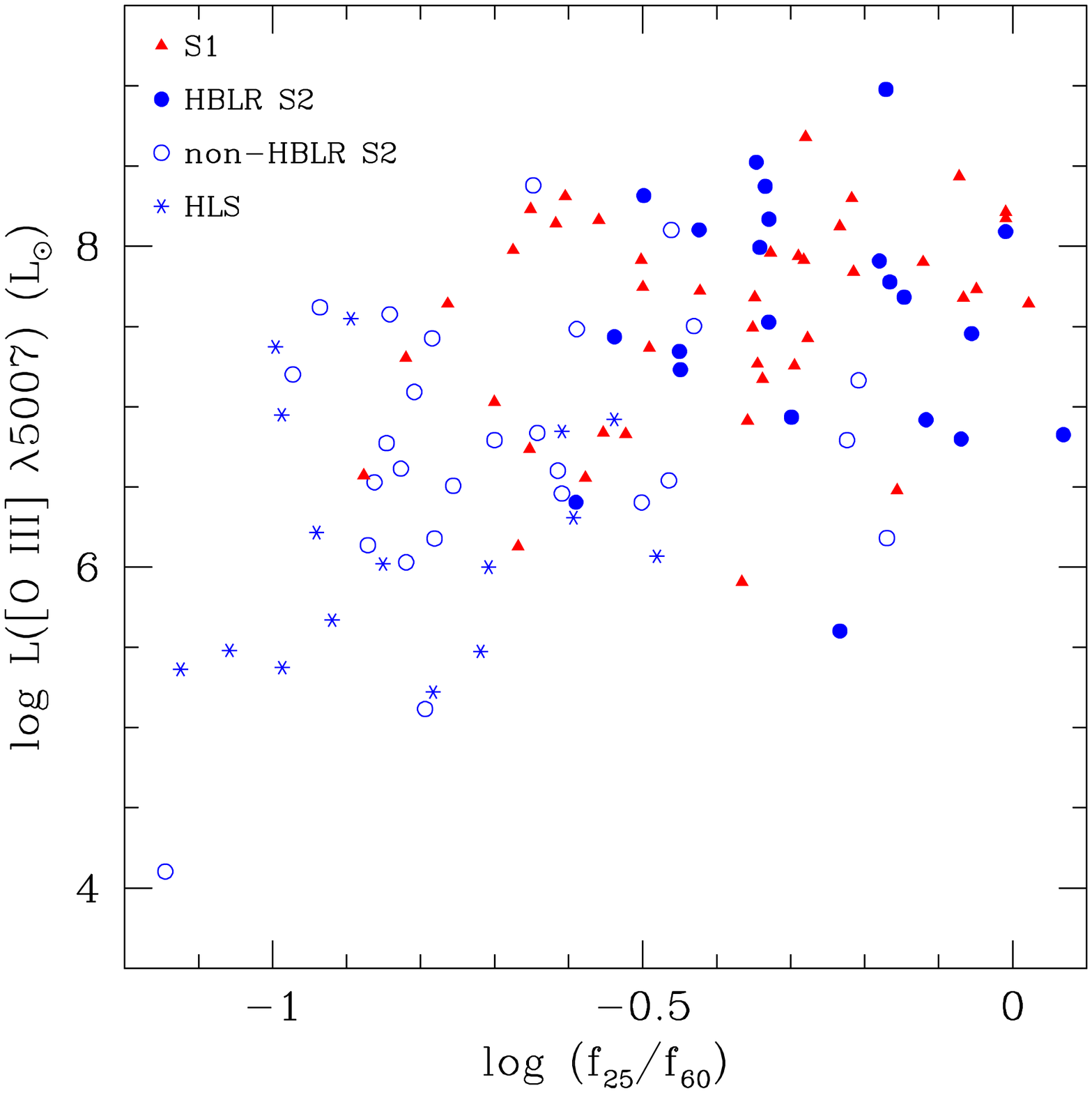,width=8.6cm}
\figcaption{\oiii~\wave 5007 luminosity versus IR color \firr~for the CfA
and \tm~samples. Symbols are as in Figure \ref{oiiihb}. Good separation 
between HBLR and non-HBLR S2s is observed in this diagram. The S1s tend 
to lie among the HBLR S2s, while largely avoiding the lower left corner,
which is occupied mainly by non-HBLR S2s and HLS galaxies.
\label{loiiiirc}}
%\vskip 0.0cm

%FIG 10
\psfig{file=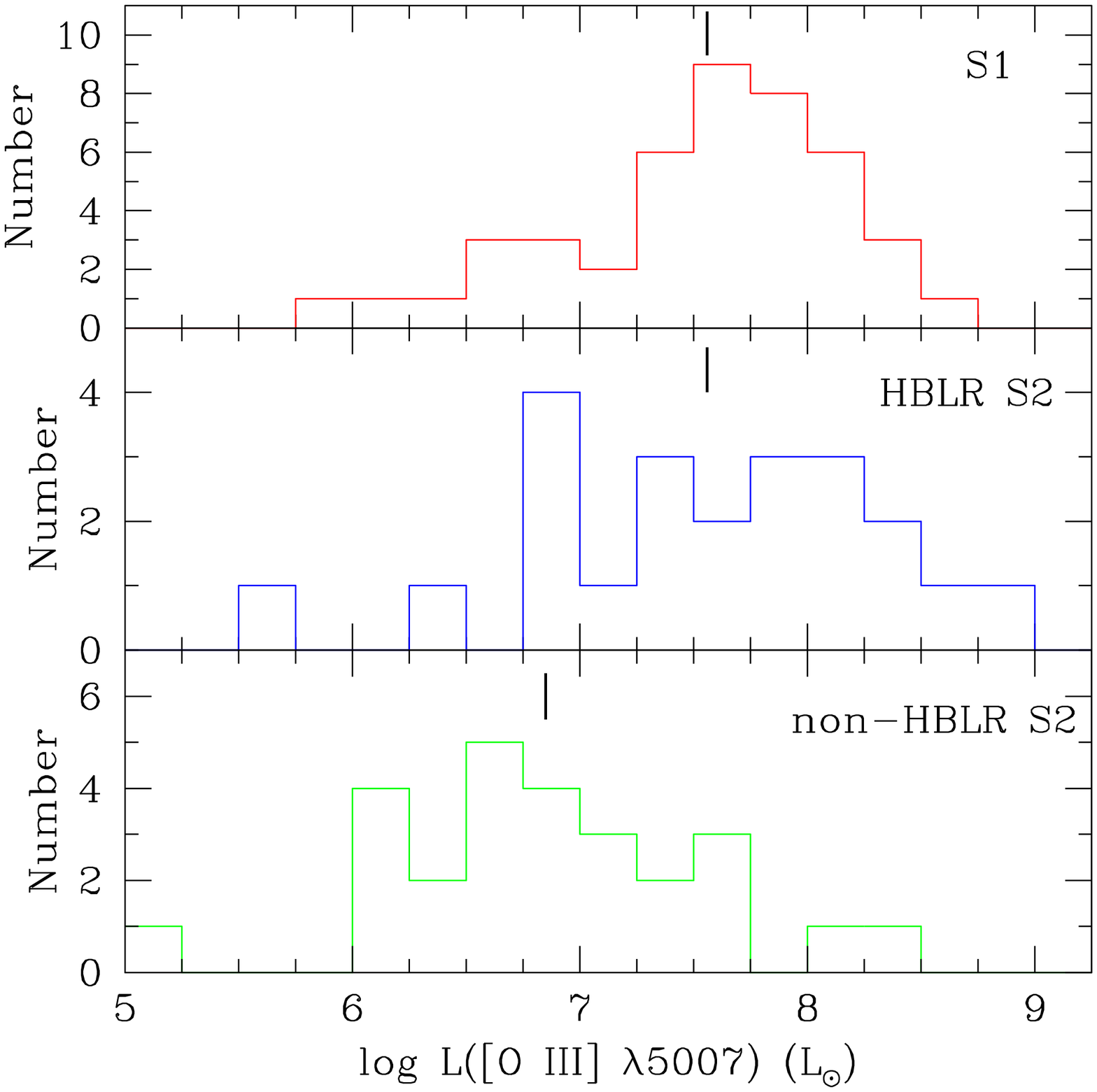,width=8.6cm}
\figcaption{The distributions of \oiii~luminosity in units of log solar
luminosity, arranged as in Figure \ref{lohbhist}. The S1s show a
similar distribution to HBLR S2s, both of which are significantly
stronger than non-HBLR S2s.
\label{loiiihist}}
\vskip 0.2cm

\noindent
weak \oiii. This diagram shows that HBLRs and non-HBLRs can be well
separated by their HX and \oiii~luminosities.

Clear separation between the two S2 types is also seen in the L(\oiii)
vs. \firr~plot, shown in Figure \ref{loiiiirc}, which is analogous to
the stellar Hertzsprung-Russell diagram. Again, the positions of S1s
in this ``AGN HR diagram'' largely overlap those of HBLR S2s but not
non-HBLR S2s.  In the accompanying Figure \ref{loiiihist}, we show the
distribution of log L(\oiii) for the three Seyfert types: S1, HBLR and
non-HBLR S2s. Here, the observed \oiii~luminosities uncorrected for
extinction are shown. As can be seen, the distributions show a
striking similarity between S1s and HBLR S2s, while there is a
significant shift to lower values for the non-HBLR S2s. This result
provides strong support for the UM in that it confirms the prediction
that isotropic properties such as L(\oiii) should be the same between
S1s and S2s, {\it but only when the HBLR S2s are considered, and
non-HBLR S2s are excluded}.  \citet{keel94} have noted the similar
L(\oiii) distributions for their sample of Seyfert galaxies selected
on the basis of far-IR flux and {\it warm} (\firr~$> 0.27$) color. We
can now understand why the S1s and S2s in their sample are well
matched in L(\oiii): since warm S2s are well-known to be largely of
the HBLR variety, non-HBLR S2s have been selected against, and thus
most if not all of the S2s in their sample are truly misdirected
S1s. This point is considered further in \S \ref{host}.  When
separation of HBLR and non-HBLR S2s is not properly performed in the
analysis, the combined sample of S2s would show a {\it smaller}
average L(\oiii) than S1s. This expectation is confirmed by our
sample, and also consistent with that implied by the results of
\citet{mr95}.

The distribution of $L_{25}$ (Fig. \ref{l25hist}) also shows a
behavior similar to that of L(\oiii), but with lower significance. It
confirms the result of \citet{la01} that HBLR S2s are more energetic
at mid-IR wavelengths than non-HBLR S2s. While the 25$\mu$m 
luminosity of S1s seems to be similar to the former, they also 
share this property with the latter.  
Finally, in terms of radio power, it has been shown (Paper I; Moran et

%FIG 11
\psfig{file=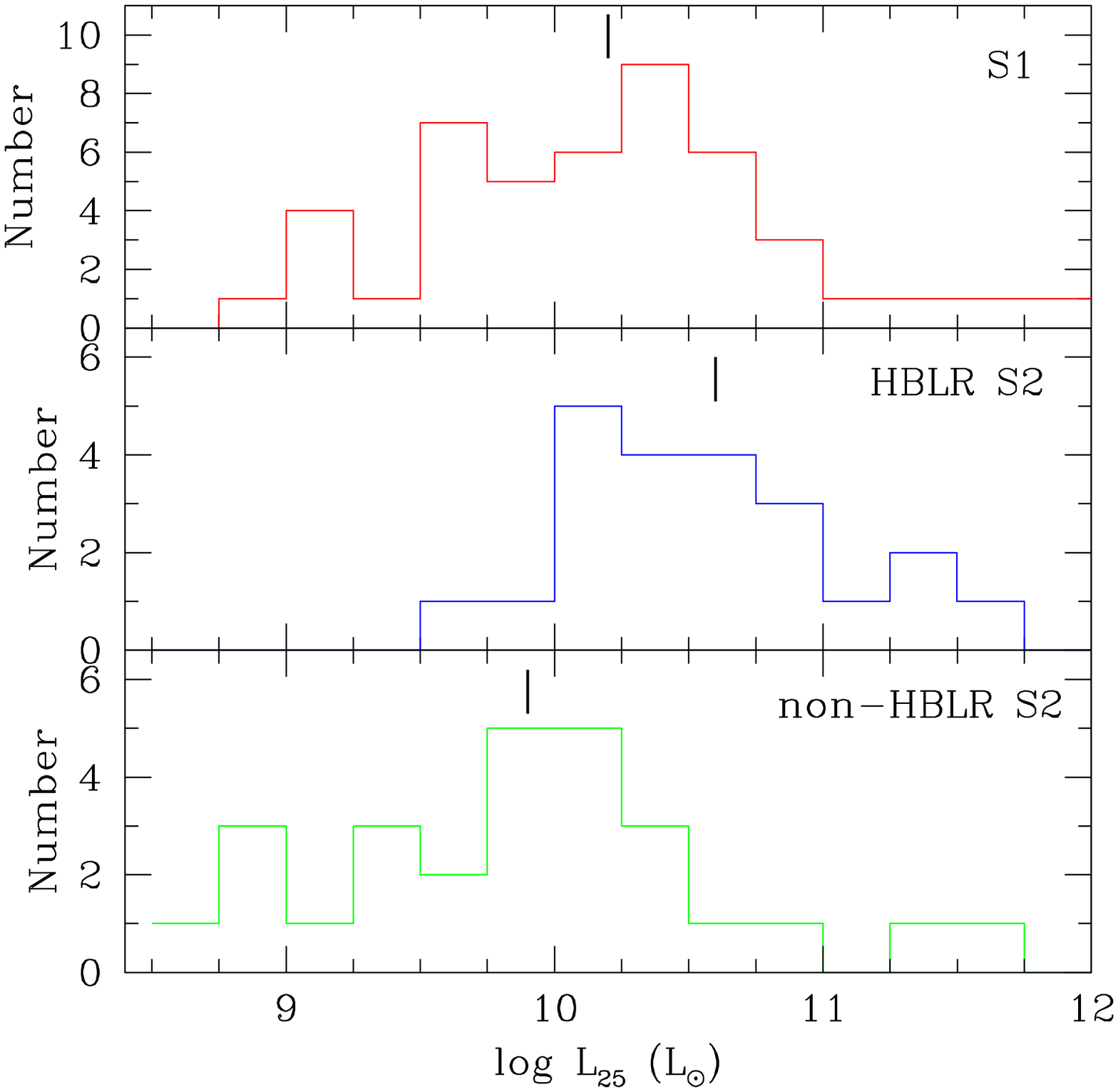,width=8.6cm}
\figcaption{The distributions of IRAS 25 $\mu$m~luminosity in units of
log solar luminosity, arranged as in Figure \ref{lohbhist}.
\label{l25hist}}
\vskip 0.2cm

\noindent
al. 1992; Thean et al. 2001) that HBLR S2s as a group are more
luminous than their weaker, non-HBLR cousins. However,
compared to S1s, HBLR S2s also appear significantly stronger, consistent with
\citet{the01}, as Figure \ref{lradhist} and Table~3 show.

Turning to the far infrared (FIR) luminosity, we find the situation to
be quite different. Following \citet{con91}, the FIR flux is
calculated according to the formula $f_{FIR} = 1.26 \times 10^{-14}
(2.58f_{60} + f_{100})$ $W~m^{-2}$.  As shown in Figure
\ref{lfirhist}, the distribution of $L_{FIR}$ is indistinguishable
among the three Seyfert types, with a mean log $L_{FIR}$ of about 10.3
\lsun. Since the FIR luminosity is a good indicator of
star forming regions (e.g.,

%FIG 12
\psfig{file=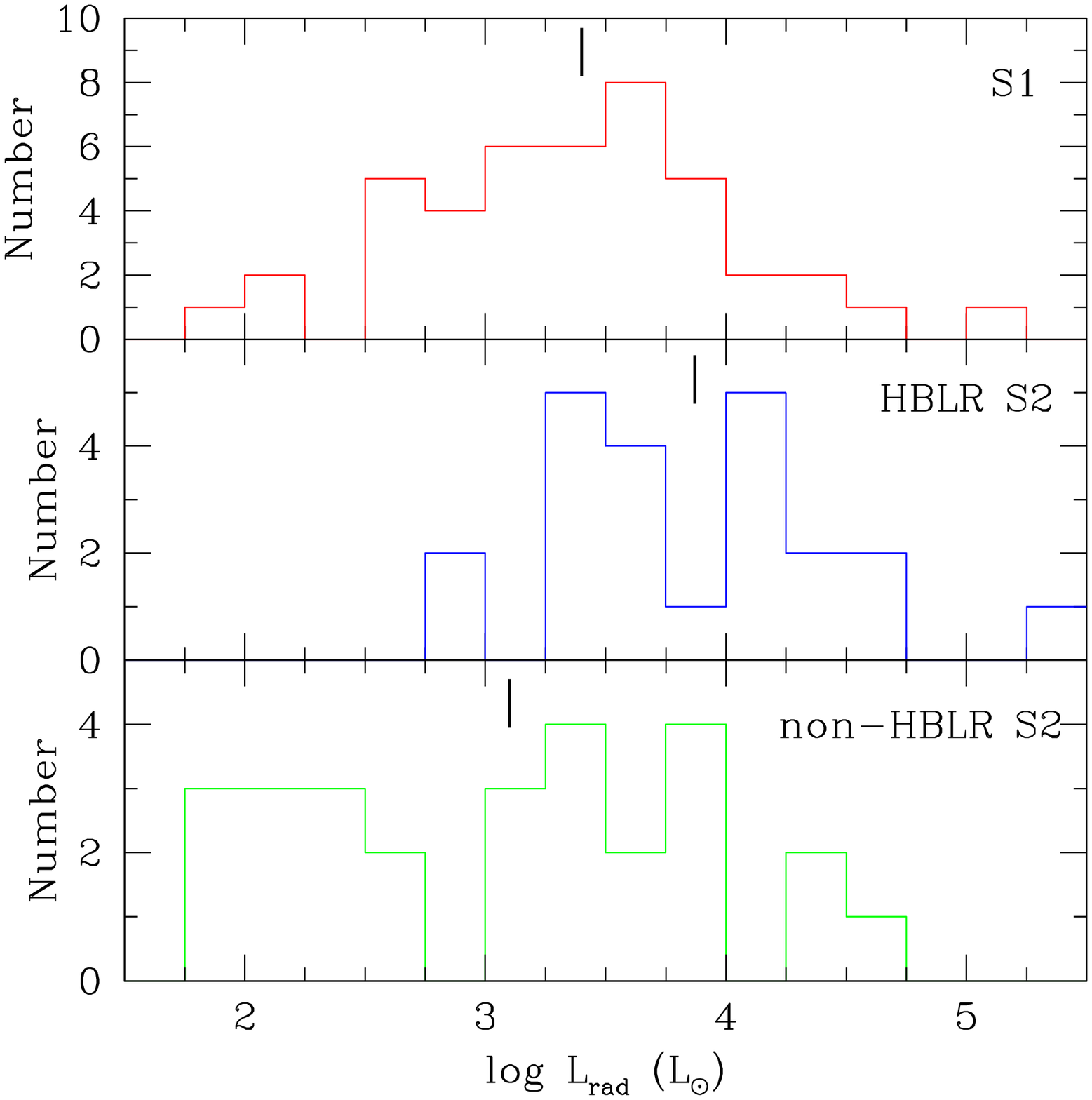,width=8.6cm}
\figcaption{The distributions of 20cm radio luminosity in units of log
solar luminosity, arranged as in Figure \ref{lohbhist}. Compared to
non-HBLR S2s, HBLR S2s are more powerful in the radio, but the latter
also appear to be more powerful than S1s. The radio luminosity has
been calculated assuming a uniform bandwidth of 45 MHz.
\label{lradhist}}
%\vskip 0.2cm

%TABLE 4
\hspace*{-0.8cm}
\scalebox{0.90}{\includegraphics[160,287][454,487]{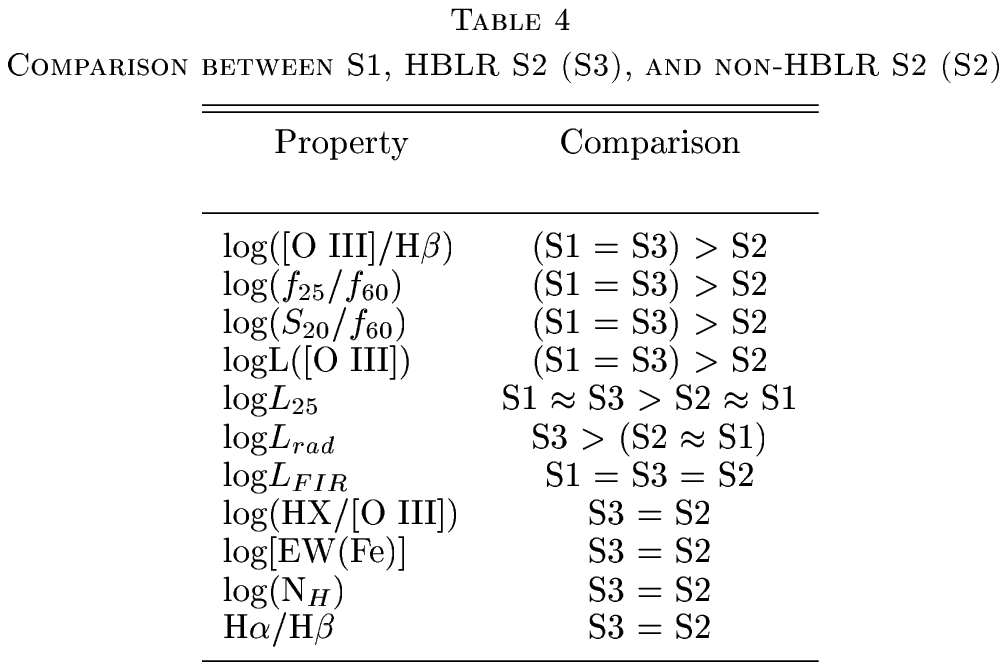}}

\noindent
Alonso-Herrero et al. 2001; Ruiz et al. 2001), this 
suggests that the circumnuclear star formation level in these classes 
of Seyferts are essentially the same. The major implication is 
that the increased AGN power observed in S1s
and HBLR S2s compared to non-HBLR S2s is due neither to the increased
obscuration nor elevated level of star formation in the latter, both
of which could effectively mask the AGN activity, but rather to the
intrinsically stronger nuclear activity of the former.  Our result is
consistent with \citet{pr01}, who found that the cold (FIR) component
in the CfA S1 and (total) S2 populations are similar. However their
finding that the warm (mid-IR) component in S1 is stronger than S2 can
be fully accounted for by the presence of non-HBLR S2s (see \S
\ref{host}).

In summary, we present in Table~4 a simple outline of the
differences and similarities in the various observational properties
discussed among the S1s, HBLR S2s and non-HBLR S2s. For simplicity, we
have denoted HBLR S2s as S3s, and non-HBLR S2s simply as S2s in the
table.  Similarity is denoted by the $=$ or $\approx$ symbols, and
significant difference is indicated by the $>$ sign.  The results show
that while obscuration and SF activity 

%FIG 13
\psfig{file=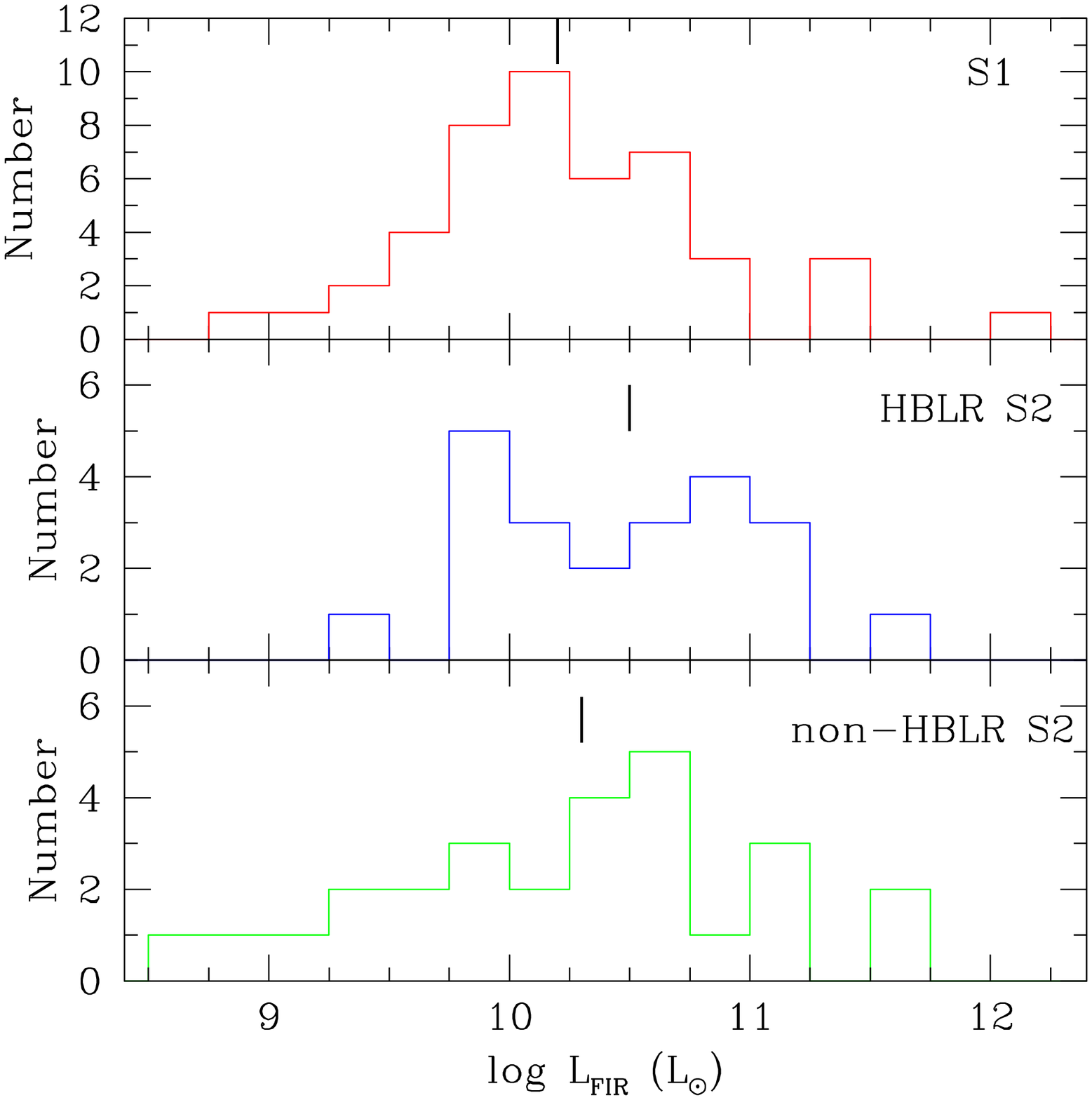,width=8.6cm}
\figcaption{The distributions of far-IR luminosity in units of log solar
luminosity, arranged as in Figure \ref{lohbhist}. The distributions
for the three classes of Seyfert galaxies are statistically identical.
\label{lfirhist}}
%\vskip 0.2cm

\noindent
seem to be similar between the two S2 types, virtually all 
measures of AGN power indicate that non-HBLR S2s are different from
and less energetic than HBLR S2s, which in turn appear to be the same
as genuine S1s. Thus, only the HBLR S2s should strictly be considered
truly S1s viewed from a different direction.

Note that even with the substantial reduction in
the true number of hidden S1 nuclei (by roughly half), the real S2
model is not inconsistent with the observed size of the ionization
cones of Seyfert galaxies.  For a typical half-opening cone angle of
$\approx$ 30\arcdeg~(e.g., Wilson \& Tsvetanov 1994), the relative
space number density of S2 to S1 is about 6.5:1. If the number of true
hidden S1s (i.e., excluding the non-HBLR S2s) is cut by half as our
data suggest, then the torus half-opening angle would rise to $\sim$
40\arcdeg. This is entirely within the range observed and fully
consistent with existing data, given the inherent difficulty and
uncertainty involved in measuring opening angles of ionization cones.

\section{Discussion} \label{disc}

\subsection{Alternatives to Two Populations of Seyfert 2s} \label{alt}

Because the interpretation of two intrinsically different S2 types
could have a potential for upsetting the currently popular UM
paradigm, it is worthwhile to examine some of the alternatives and make
certain that the two kinds of S2s are genuinely different.  Besides
being intrinsically weaker, possible reasons that the non-HBLR S2s may
in fact be normal obscured S1s, but are somehow able to escape
detection include: 1) S/N in some spectropolarimetric observations may
simply be too low to detect weak HBLRs.  This problem is exacerbated
by the limited resolution (seeing) of ground-based telescopes, which
must extract and detect precious scattered (polarized) photons against
an overwhelming background of unpolarized starlight and 2) Placement
and orientation of the spectroscopic slit in these observations may
have missed a small, or well-collimated scattering region. However, 
while some HBLRs could have escaped the detection limit of the survey,
these possibilities alone cannot explain why the two S2 types lie in such
separate regions of the diagnostic diagrams discussed above, and having very
distinct luminosity distributions (Figs. \ref{loiiihist}--\ref{lradhist}),
with the HBLR S2s being generally more aligned with the S1 population.

Since the detectability of broad polarized \ha~scales with the
strength of the emission line, we can assess the detection limit of
HBLR in our survey by examining the distribution of the observed
fluxes for some of the best indicators of AGN strength.  Shown in
Figure \ref{foirrad} is the plot of the observed extinction-corrected
\oiii~fluxes against the 25$\mu$m mid-IR and 20cm radio flux
densities. A striking characteristic of this figure is that, contrary
to the very significant differences seen in {\it luminosity} space (\S
\ref{diag}) for these three properties, there is no separation at all
in flux space between the two S2 types. K-S tests show that these flux
distributions are virtually the same between HBLR and non-HBLR S2s,
with $p_{null}$ ranging from 11\% to 44\%. Thus a standard BLR is not
any more likely to get detected in an HBLR S2 than a non-HBLR S2. It
is also clear from Figure \ref{foirrad} that many HBLR detections
reach to very low observed flux level, below those of many non-HBLR
S2s. For example, some of the HBLR detections are as much as an order
of magnitude below the detection limit of $f_{[O III]} = 10^{-12}$
ergs s$^{-1}$ cm$^{-2}$ estimated by \citet{al01} for a 4-m class
telescope. Thus, we conclude that there is no strong evidence for an
observational bias against the detection of HBLRs in non-HBLR S2s, and

%FIG 14
\psfig{file=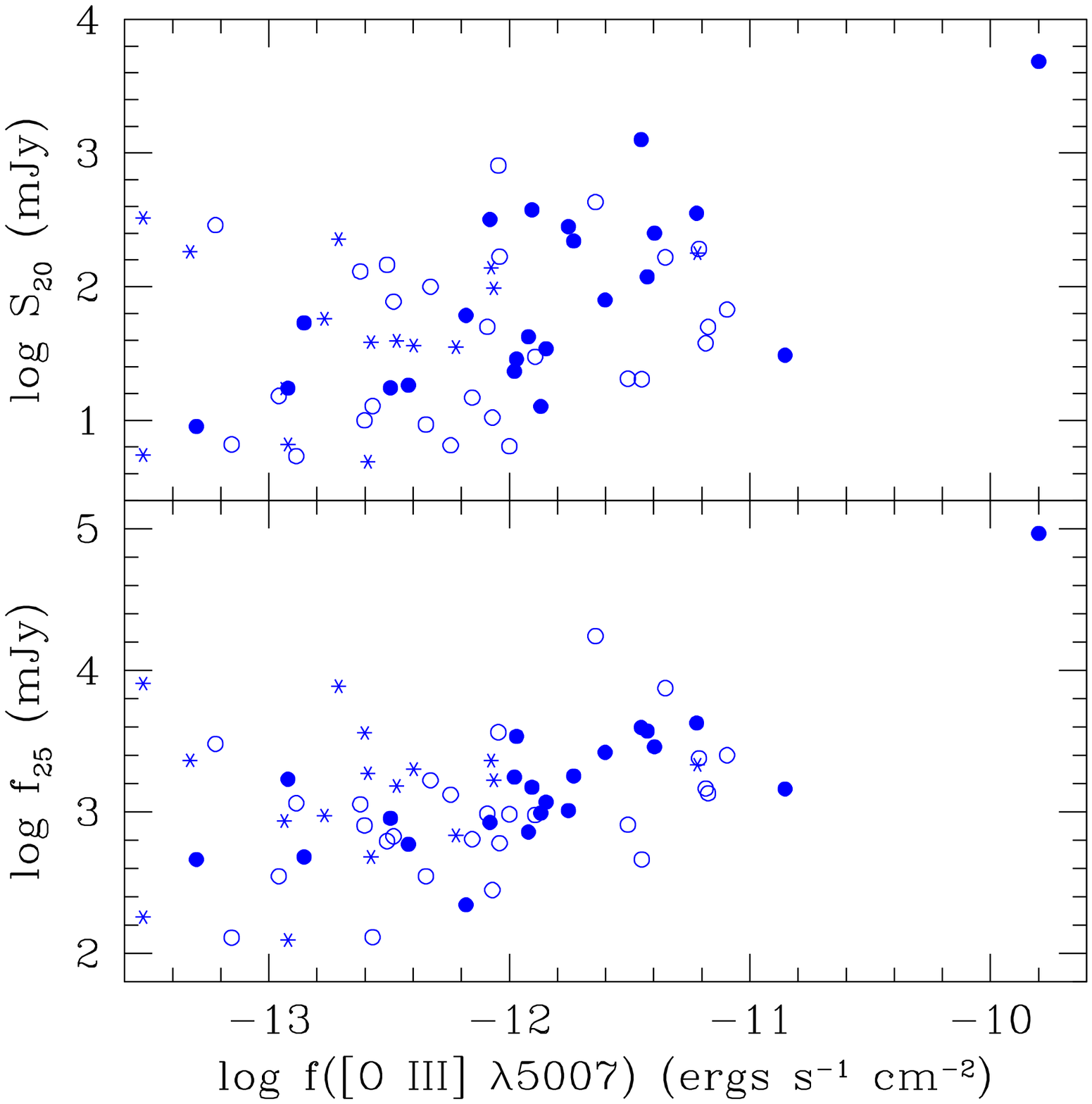,width=9.0cm,height=9.2cm}
\figcaption{Extinction-corrected \oiii~\wave 5007 fluxes, 25$\mu$m mid-IR, 
and 20cm radio flux densities for the S2 and HLS galaxies in the CfA 
and \tm~samples. These three properties have been shown to be a good 
indicator of the AGN strength. Symbols are as in Figure \ref{oiiihb}. 
The upper-rightmost data point refers to NGC 1068. HBLR detection can reach
to very low observed flux level, below those of many non-HBLR S2s. 
In addition, the flux distributions are the same for the two S2 types,
indicating that the non-detections of HBLRs cannot simply all be due to
the detection limit of the survey. 
\label{foirrad}}
\vskip 0.2cm

\noindent
their non-detections cannot simply all be due to the detection limit of the survey.

Paper I also addressed the question of whether the non-detection
of HBLRs in non-HBLR S2 is due to the lack of an energetic AGN (and hence
BLR), or to such exceptionally high obscuration that no signal from
the buried AGN could be detected (i.e., $S_{20cm}$ and \ftv).
\citet{mg90} and later \citet{hlb97} suggested that the scattering may
take place very close to the nucleus, in the inner ``throat'' of the
torus, and non-HBLR S2s are perhaps those with the torus axes tipped
at larger inclinations, resulting in higher obscuration and greater
obstruction of the scattering region. They would be expected to
represent the Compton-thick AGNs, which show the highest X-ray
absorption with $N_H$ in the range $10^{23-24}$ cm$^{-2}$ and beyond.
Aside from being in contrast with the observations that very extended
scattering regions have been seen in many AGNs (e.g., Miller,
Goodrich, \& Mathews 1991; Tran et al. 1998, 2000; Cohen et al. 1999)
this picture cannot be correct, as
there are hints that very large inclinations have indeed been seen in
HBLR AGNs \citep{t99}. In addition, as found by \citet{al01} and Paper
I, and discussed in the previous section, there is essentially no
significant difference in the hard X-ray column density or other
obscuration indicators, such as HX/\oiii~and EW(Fe), between the HBLR and
non-HBLR S2s, contrary to expectation if indeed HBLR S2s are
preferentially viewed more pole-on. Another difficulty for the model is
that mid-IR arguments indicate that there is {\it no} apparent
difference in the optical depth between S1s and HBLR S2s, while 
that between the two S2 subtypes is unreasonably high (see \S \ref{host}).
Paper I also found that the Balmer decrement between the two S2 subtypes
are similar (see Table~3), and Lumsden et al. (2001) have 
dismissed the notion of \firr~being an indicator of viewing angle.

As emphasized in Paper I, the lack of HBLR in the non-detected objects
also cannot be easily attributed to a ``contrast effect'', where the
overwhelming contribution of starlight in the host galaxies may render
any polarized signal difficult or impossible to detect. Starlight
level in HBLR S2s reaching $\sim$ 80-90\% is quite common
\citep{t95,bfm99}. Both HBLR and non-HBLR display similar levels of
starlight domination, and the non-detection of a HBLR seems to be
unrelated to it (Kay \& Moran 1998; Tran et al. 1999), 
nor can it be simply attributed to a strong starburst coexisting with the AGN,
``muddling'' the picture. In this scenario, non-HBLRs may be cases where
the starburst component is unusually strong, capable of contributing
substantially to, and ``contaminate'' the total energy output (e.g.,
Levenson et al. 2001b). 
\citet{ghl01} examined the question of whether starburst-dominated
(i.e., ``composites'') and non-HBLR S2s are the same class of objects,
using literature data available prior to our survey.  They concluded
that the answer appeared to be no.  With the present extended dataset
we confirm that {\it not all} starburst-dominated S2s are non-HBLR S2s
and vice versa. Many starburst-dominated S2s have been
found to be HBLRs but, by definition, {\it none} of the non-HBLRs are
HBLRs. For example, there are 11 HBLR S2s detected to date in the
sample of \citet{cid01}, with about half belonging to either the
starburst/S2 composites or the normal S2s. So the relationship between
starburst-dominated composites and non-HBLR S2s is not simple.
Thus, the occurrence of HBLR shows no strong preference for either the
``composite'' or ``pure'' systems. In addition, the average FIR luminosity of
the two S2 types are indistinguishable, indicating no difference in
their star forming or bursting properties.

Another possibility is that the nature of the obscuring medium may
influence the detection of HBLRs.  The non-HBLR S2s may belong to a
class of AGN where the obscuring medium isn't a torus at all, but may
take a form of a much more extended interstellar medium (MGT98) in the host galaxies.
In this case, no strongly collimated ``scattering cones'' are
expected, and thus no polarized broad lines are observed.  In this model, the
non-HBLR S2s would then represent largely those that show little or
only modest X-ray absorption, with column densities expected to be in
the range of Galactic values (i.e., $N_H \sim 10^{20-21}$, or
Compton-thin).  Again, this is contrary to their observed
distributions of obscuration indicators discussed earlier.
Alternatively, the torus opening may not be the same for all galaxies,
but is variable in size (e.g., Lawrence 1991), becoming larger for
increasing AGN luminosity.  In this scenario, the BLR and obscuring
torus do exist in non-HBLR S2s with properties similar to those with
detectable HBLR (i.e., similar dust to gas ratio, composition, $N_H$,
etc.) but the torus cone angles are considerably narrower, so narrow
in fact that little or no ionizing radiation, and hence reflected
light, could escape and be detected. One difficulty with this
interpretation is that some of the most spectacular cases of
ionization cones, which are neither narrow nor lacking in ionizing
radiation, are found in non-HBLR S2s: Mrk573, NGC5728 \citep{wt94,sk96}.

If a hidden S1 exists in non-HBLR S2s, the simplest explanation for
its non-detection may be that it is too weak to be detected, and thus
observationally ``lacks'' a BLR.  The scattering region simply may not
be able to exist in a Seyfert galaxy hosting an intrinsically weak S1
nucleus, or it may be too small to enable sufficient flux to be
scattered \citep{la01}, and thus allowing HBLR to be more easily
detected. Again, the lack of HBLR in sources with spectacular
ionization cones is puzzling.  Although none of the alternative models
discussed seems to satisfactorily explain the differences and
similarities presented in the previous section amongst the S1s and
HBLR and non-HBLR S2s, this last interpretation is not inconsistent
with the concept of two S2 populations: a more powerful (with HBLR)
nucleus would be expected to support a larger scattering region, while
a weaker (non-HBLR) one may only be able to sustain a much smaller one
or not at all. Similarly, the receding torus hypothesis can also accommodate
the changing AGN strength with the varying opening of the torus that is 
central to the UM, potentially reconciling the two ideas.

\subsection{Large-scale Properties of Seyfert Galaxies} \label{host}

Given the seeming dichotomy of S2s suggested by our study, it would be
of interest to ask if many of the differences found between S1s and
S2s in the past could be explained by the fact that this dichotomy had
not been taken into account.  When data from previous studies, which
assumed that {\it all} Seyfert galaxies of type 2 were equivalent, are
reanalyzed in terms of the two S2 populations, with one being truly hidden
S1s and the other real S2s, do the large-scale differences found in previous
studies tend to disappear? This is the question that we would like to explore 
in this section. 

There is already a hint that this is in fact the case when we examine
the study of \citet{s01}, which selects Seyfert galaxies based on
far-IR flux {\it and warm} color. Unlike numerous earlier
investigations, this study found no statistically significant
differences between S1s and S2s in various properties such as, the
host galaxy morphologies and frequency of companion galaxies.  We
argue that the differences reported in the past between S1s and S2s
(e.g., MGT98; Dultzin-Hacyan et al. 1999) may have gone away in the
Schmitt et al. study not, as has been claimed, because this sample is
any more complete and based on more isotropic property than previous
surveys, but {\it precisely} because of its selection effect: the
sample preselects only {\it warm} Seyferts (\firr~$> 0.27$),
effectively discarding all weaker non-HBLR S2s. In other words, this
sample really compares between the normal S1s and their truly hidden
counterparts: warm, HBLR S2s. Once this warm criterion is relaxed, 
the sample would undoubtedly contain substantial number of weak or
``real'' S2s, LINERs, starbursts, and HII galaxies, all of which have
a strong dust (far-IR) component. Similar differences that were found
in previous studies would likely be present again. A check on the
statistics of HBLRs in the Schmitt et al. sample shows that about 70\%
of its S2s harbour HBLRs. The data are not complete (only $\sim$40\% of
the S2s in the sample have currently been observed
spectropolarimetrically), but this HBLR frequency is substantially
higher than what was found by all previous surveys (Paper I; Moran et
al. 2000; Lumsden et al. 2001).  On the other hand, the sample of
\citet{dr98} study, in which very significant difference is found
between the mean environment of S1s and S2s, contains only one known
HBLR S2 (NGC 4388). This study, therefore, compares properties mostly
between S1s and non-HBLR S2s.  These indications strongly argue for
the concept that the two types of S2s are fundamentally different in
nature. They also underscore the importance of separating out the
truly hidden S1s (i.e., HBLR S2s) from the ``real'' S2s when comparing
their properties to normal S1s.

To further illustrate that there may be no real difference between S1s
and S2s when the HBLR and non-HBLR subtypes are properly accounted
for, we will take advantage of the marked increase in the currently
known HBLR S2 population as a result of several recent
spectropolarimetric surveys and reexamine some of the data of previous
studies.  We only consider HBLR S2s as the truly obscured S1 galaxies
while excluding non-HBLR S2s from the comparison.  When this is
performed, the differences between S1s and S2s found by these studies
tend to be insignificant. This provides one of the most compelling evidence
to date for two S2 populations. We note that although some of our
results may suffer from small-number statistics and/or incomplete
samples, they are nevertheless useful as consistency checks of our
hypothesis.  Our reexamination includes observational evidence from
the following studies:

1) Clavel et al. (2000) show that essentially all the S2s known to have
HBLRs display mid-IR $Infrared~Space~Observatory$ ($ISO$) spectra that look just like S1s, but those of
S2s without HBLRs are indistinguishable from starbursts. We find
that the equivalent widths (EWs) of the 7.7$\mu$m feature, usually attributed 
to polycyclic aromatic hydrocarbon (PAH), and the
underlying local continuum luminosity show significant differences between 
HBLR and non-HBLR S2s. The Clavel et al. sample contains five HBLR S2s and
10 non-HBLR S2s that could be identified. The mean 7.7 $\mu$m PAH EWs
are 0.921 $\pm$ 0.535 $\mu$m and 3.59 $\pm$ 1.72 $\mu$m for HBLRs and 
non-HBLRs, respectively. The corresponding means for the 7$\mu$m 
monochromatic continuum luminosity are 
$\langle {\rm log}\nu L_{\nu,7} \rangle = 43.7 \pm 0.61$ ergs s$^{-1}$ 
and 42.6 $\pm$ 0.51 ergs s$^{-1}$. These distributions are 
different at the 0.9\% and 2.8\% significance level, respectively. 
On the other hand, their mean 7.7$\mu$m PAH {\it luminosities} are 
indistinguishable at the 66\% significance level 
(HBLRs S2s: $\langle {\rm log}L_{7.7} \rangle = 42.7 \pm 0.74$ ergs s$^{-1}$;
non-HBLRs S2s: $\langle {\rm log}L_{7.7} \rangle = 42.2 \pm 0.65$ ergs s$^{-1}$). 

Interestingly, the same quantities for S1s 
($\langle EW_{7.7} \rangle = 0.53 \pm 0.47$;
$\langle {\rm log}\nu L_{\nu,7} \rangle = 43.73 \pm 0.85$;
$\langle {\rm log}L_{7.7} \rangle = 42.44 \pm 0.80$)
display nearly identical behaviors compared to 
HBLR S2s, but not to their non-HBLR counterparts: 
both S1 and HBLR S2s show much higher 7$\mu$m continuum luminosity
and lower 7.7$\mu$m PAH EWs than non-HBLR S2s,
while their 7.7$\mu$m PAH luminosities are about the same. 
Since the PAH features are generally associated with intense starbursting 
regions, photo-dissociation regions and galactic cirrus on a much larger scale 
unrelated to the nuclear activity (e.g., Laurent et al. 2000), 
these indications again suggest that the level of star formation is similar 
in these Seyfert galaxies; it is the active nuclear engine that is different. 
The differences and similarities found by \citet{clav00} between the two
main classes of S1s and S2s, therefore, can entirely be attributed to the 
presence of non-HBLR S2s in their sample. They interpreted these differences 
as being due to orientation effects similar to those proposed by \cite{hlb97}. 
As argued in \S \ref{alt}, however, this model is untenable, and thus cannot 
properly explain them. Specifically, \citet{clav00} used the PAH EW
as an indicator of the nuclear obscuration and derived an average difference
in visual extinction of $A_V \approx 92$ mag between S1s and S2s. 
%in column density of $N_H = 2\times 10^{23} cm^{-2}$ between S1s and S2s.
However, the same analysis would indicate a similar and unreasonably large
difference in obscuration between the non-HBLR and HBLR S2s, and virtually
{\it no} difference between HBLR S2s and S1s, contrary to the UM. 
Rather than a reddening indicator, the PAH EW should more appropriately 
be viewed as a measure of the intrinsic nuclear {\it strength}. The mid-IR
radiation therefore, is a good measure of AGN activity.  

2) From the three-component modelling of the ISO spectra of CfA Seyfert 
galaxies by \citet{pr01}, our examination shows that HBLR S2s have a strong 
warm dust component similar to that found in S1s, while non-HBLR S2s are  
characterized by dust that is generally cooler. The mean $F_{warm}/F_{IR}$
ratio for the four HBLR S2s in the \citet{pr01} sample is 0.45 $\pm$ 0.13. 
This is comparable to the value of 0.42 for S1s, but much higher than 0.25
found for the non-HBLR S2s. Similarly, the mean $F_{warm}/F_{20cm}$ for 
HBLR S2s is 6.56 $\pm$ 0.34, comparable to 6.5 for S1s, but higher than 6.1
for other S2s \citep{pr01}.  

From their observations, both \citet{clav00} and \citet{pr01} have
indicated that the obscuring torus, if it exists, cannot be as
optically thick as had been thought (e.g., Pier \& Krolik 1992).  As
also suggested by other studies, even at near to mid-IR wavelengths,
the AGN radiation appears to be isotropic \citep{gdf97,fad98}, and may
suffer from less extinction than commonly thought
\citep{vgh97,ris00,ah01}.  This is confirmed by our finding that the
distributions of $L_{25}$ for S1s and S2s are very similar.
(Fig. \ref{l25hist} \& Table~3). As also demonstrated in
\S \ref{diag}, Paper I, and Lumsden \& Alexander (2001), the
well-known warmer \firr~ratio in the HBLRs S2s compared to non-HBLR
S2s is essentially a result of the former being intrinsically more
luminous in mid-IR, and suggests that even for highly obscured AGN,
the mid-IR signature of the powerful AGN can be seen. A lower optical
thickness would also be consistent with evidence for 
lower optical/IR extinction than expected from hard X-ray column density, 
due perhaps to larger grain size in AGNs (e.g., Maiolino et
al. 2001a,b; Imanishi 2001), or the different spatial regions probed by the 
two wavelength regimes \citep{ris00,wm02}. For this reason, the enormous $A_V$
values often deduced from the hard X-ray $N_H$, assuming standard
dust/gas and extinction curve, are highly suspect.

3) Based on a $HST$ snapshot imaging survey of a large, but heterogeneous 
sample of Seyfert galaxies, MGT98 found that the large scale environment of
Seyfert 1 and 2 galaxies are significantly dissimilar in terms of
their dust morphologies.  However, when the Seyfert sample of MGT98
are grouped separately into HBLR and non-HBLR S2s, the dust
morphologies of the host galaxies of the former are statistically the
same as S1s, which in turn, are different compared to non-HBLR S2s.
In the MGT98 sample, the fraction of non-HBLR S2s that shows either
dust lanes or absorption patches (designated D, DC, DI in their paper)
is 10/18, or 55\%.  The corresponding fraction for HBLR S2s is 3/11
(27\%). Not only is this significantly lower than that for their
non-HBLR cousins, it is essentially the same as for S1s
(23\%). Moreover, given that the dust incidence was found to be 39\%
for the total S2 population (MGT98), the above fractions are perfectly
consistent with our finding that about half of them belong to each of
the non-HBLR and HBLR subclass [i.e., $(27 + 55)/2 = 41$\%].

4) As mentioned in this study, HBLR S2s have \firr~and
[OIII]/\hb~similar to S1s (see Fig. \ref{oiiihb}), while in non-HBLR
S2s these ratios tend to be significantly smaller. Although the
\oiii/\hb~ratio is on average smaller in non-HBLR S2, it is still well
above the canonical value of 3 for Seyfert galaxies. Thus these are
truly bona fide Seyfert galaxies, not mis-classified HLS objects, and
the possibility that the latter may have ``contaminated'' the S2
sample has been eliminated.  Rather, it is more likely that the line
ratio can be explained by fundamental difference in nuclear
strength. This is further reinforced by the discovery that isotropic
properties such as, L(\oiii), and $L_{25}$ are found to be
statistically the same between S1s and HBLR S2s, but they are
significantly lower in non-HBLR S2s than in their HBLR counterparts
(\S \ref{diag}).

\citet{s98} also compared several emission line ratios between S1s and
S2s in his study.  He found that the \oii/\neiii~and \oii/\nev~ratios
are statistically lower in S1s compared to S2s, indicating a higher
excitation spectrum in the former. However, when a similar comparison
is made between S1s and HBLR S2s only, we find that these differences
are no longer statistically significant. The mean \oii/\neiii~ratio for
the 12 HBLR S2s found in his sample is 2.6 $\pm$ 1.4. A K-S test
against the S1 sample distribution (with mean 1.73 $\pm$ 0.8) yields
$p_{null} =$ 23\%, confirming their similarity.  For the
\oii/\nev~ratio, the number of available data for HBLR S2s is
considerably smaller, rendering a statistical test less accurate, but
it appears that there is also no significant difference between the
mean of five HBLR S2s (2.3 $\pm$ 1.6) and that of S1s (1.54 $\pm$
1.6).  Thus, combined with the \oiii/\hb~property discussed earlier,
these emission line characteristics strongly suggest that the
ionization of the narrow line regions are very similar between S1s and
HBLR S2s, while in non-HBLR S2s, it is statistically weaker.  The
puzzling line-ratio differences found by \citet{s98} between the two
main Seyfert types need not invoke a special alignment of the torus
axis with the host plane axis in S1s as had been proposed. They can
instead simply be explained by the existence of two populations of
S2s, only one of which truly contains genuinely powerful hidden S1
nuclei capable of fully ionizing the extended NLR.

\subsection{Evolutionary Sequence of Seyfert Galaxies} \label{evol} 

The possibility of two types of S2s has enormous implications for the
nature of the Seyfert phenomenon and unification model of AGN. For
example, it shows that orientation alone does not fully account for
the differences seen in all S1s and S2s.  Moreover, it suggests that
the fraction of HBLR should increase with AGN power as determined, for
example, by the radio luminosity of the AGN. This appears to be borne
out by existing observations.

Let us assume that the fraction of HBLR detected corresponds directly
to the fraction of true AGN in the population. By ``true'' AGN, we
mean energetic processes dominated by accretion power from a
supermassive black hole, {\it and} the existence of a detectable BLR.
\citet{c99} found that when combined with the results of Hill,
Goodrich, \& DePoy (1996), forming a complete, volume-limited sample
of powerful narrow-line radio galaxies, the fraction of HBLR detected
is 6 out of 9 (or 67\%). In the radio-weak LINERS, the fraction of
broad-line AGN is comparable to or less than that found in the CfA S2
sample: of those HLS galaxies observed in the \tm~sample we find
virtually none; \citet{bfm99} found 3/14 (or 21\%) in a random sample
of LINERs.  A systematic trend is noted in these surveys: the higher
the radio power of the objects, the higher the fraction of AGN found
to possess HBLRs.  This progression mirrors a similar trend already
noted in the ULIRGs: the higher the IR luminosity, the higher the
fraction of true AGNs found in the sample (e.g., Veilleux
\etal~1995). This provides support for the receding torus model of AGN
\citep{law91}, in which the torus opening increases with AGN
luminosity.  It also implies that low-luminosity AGNs, such as
LINERs/Seyferts may undergo an evolutionary process, already
implicated in the higher luminosity ULIRGs and QSOs, in which nuclear
activity is triggered, most likely through interactions with nearby
neighbors, creating starbursts, (re)fueling the central massive black
holes, and eventually forming AGN nuclei with BLRs
\citep{o93,h95,v01}. In this scenario, the real S2s and hidden S1s may
simply be at different stages of this evolutionary path (see Hunt \&
Malkan 1999).  A truly active nucleus with BLR may arise once the
activity level has reached above a threshold (e.g., Nicastro 2000), a
notion also implied by the recent radio study of Ulvestad \& Ho
(2001), who suggested that a minimum level of activity is required for
the Seyfert radio source to break out of its central engine.

That the non-HBLR S2s display emission line ratios that qualify them
as genuine Seyfert galaxies requires that there be some sort of hard
nonstellar ionizing continuum.  As shown in \S \ref{diag} and
\ref{host}, their spectra are generally characterized by lower
excitation and lower luminosities.  Any ``contamination'' by
circumstellar starbursts may contribute to the lower ionization level
observed in these objects, but not their overall \oiii, mid-IR and
radio luminosities. Thus not all non-HBLRs are necessarily composites
or have strong starburst components. As already discussed, many composites 
and starburst-dominated sources have also been found to possess HBLRs.
The alternative is that the non-HBLRs are simply intrinsically
weaker. Perhaps the central BH is less massive, or the accretion rate
is smaller in these objects.  The existence of a black hole mass --
radio power relationship \citep{fran98,mcl99,gcj01,ho01,wh01}, and the
possible correlation between the incidence of broad-line objects with
radio power strongly suggest that the BH mass could play a
crucial role in the AGN strength.  Nicastro (2000) also suggested that
there is an accretion rate threshold above which BLR would appear.
Thus, while it appears that much of the difference between S1s and S2s
can be explained solely by orientation, it would be difficult for
the same model to apply among the HBLR and non-HBLR S2s without
invoking intrinsic physical differences. Again, it is
reasonable that there is a component of evolution in this, that the
non-HBLRs may represent dormant, ``low-state'' S2s, whose activity has
yet been fully triggered. An evolutionary proposal to explain the
starburst-Seyfert connection, which could be related to the
development of the strength of the AGN engine, and hence its BLR property,
has also been envisioned by \citet{sb01}, \citet{cid01}, and Krongold, 
Dultzin-Hacyan, \& Marziani (2002).
Alternatively, real S2s may simply be those that have exhausted their fuel, 
and may not be directly connected to S1s strictly through evolution. 

\section{Conclusions}

We present evidence supporting the view that HBLR S2s are
intrinsically more powerful than non-HBLR S2s. The positive detection
of BLR in HBLR S2s appears to be due largely to the intrinsic strength
of the hidden AGN nucleus rather than the lower level of obscuration
or reduced dominance of circumnuclear starburst. When the intrinsic
difference between HBLR and non-HBLR S2s is taken into account, it is
shown that the former, on average, share many similar large-scale
characteristics with S1s, as would be expected if the UM is correct,
while the latter do not. These results strongly suggest that not all
S2s are intrinsically similar in nature, and HBLR S2s may be the
only true counterparts to normal S1s. The incidence of HBLR is found
to have a tendency to increase with AGN strength, suggesting a
temporal development in the torus opening angle, perhaps as the
nucleus evolves from a state of relative quiescence to full-scale AGN
engine.

While our findings suggest two separate types of S2s and their
evolutionary connection to S1s and each other, our study may suffer 
from selection effects inherent in samples not selected by isotropic 
properties (e.g., see Ho \& Ulvestad 2001), such as those of the CfA
and \tm~samples. Small-number statistics and limited survey depth may also
complicate some of the results. Future, deeper study of a more complete,
unbiased sample of Seyfert galaxies will provide a firmer picture,
and further test the idea proposed in this paper. 

\acknowledgments 
I thank M. H. Cohen, J. S. Miller, W. van Breugel and
H. Ford for their support while this research was being carried out
through the years at Caltech, Lick Observatory, LLNL, and the Johns
Hopkins University.  I am grateful to the staff at Lick, Palomar and
Keck Observatories for their expert assistance during the
observations, and to G. Harper, P. Ogle, and R. Vermeulen for
assistance with some of the observations and data reduction. I also
wish to thank R. Antonucci, M. Malkan, T. Heckman, J. Krolik and N. Levenson 
for useful discussions. Support from the NASA ACS grant and the Center
for Astrophysical Sciences at JHU is gratefully acknowledged.
%We acknowledge very helpful and constructive comments from the second referee,
%but do not appreciate destructive ones from the first.
W. M. Keck Observatory is operated as a scientific partnership between
the California Institute of Technology and the University of
California, made possible by the generous financial support of the
W. M. Keck Foundation.  Work performed at the Lawrence Livermore
National Laboratory is supported by the DOE under contract
W7405-ENG-48.  This research has made use of the CATS database
\citep{v97} of the Special Astrophysical Observatory, and the
NASA/IPAC Extragalactic Database (NED), which is operated by the Jet
Propulsion Laboratory, California Institute of Technology, under
contract with NASA.

%\clearpage

\end{document}